# On the formation of metallic glass

Wang Jian Guo

College of Mechanical Engineering, Dongguan University of Technology, Dongguan 523808, China

## Abstract

The high cooling rate needed for preparing the metallic glass (MG) makes the nonequilibrium nature of glass formation more prominent and requires a better quenching technique than ever before. Here, we formulate the cooling process in an analytical way and figure out the determinants for cooling rate, and analyze the crystallization time with consideration of phase diagram. Based on the reduced glass transition temperature, $T_{rg}$, for measuring the glass-forming ability (GFA), a more reasonable $\Delta T_{rg}$ is proposed. Glass transition, especially in ever glass whose ground state is of glass, is discussed in terms of thermodynamics for phase transition. A fundamental law concerning the changing rate of entropy in a closed system is supposed to underlie the physics for glass formation. These results may help understand the glass formation principally and develop new and robust MGs technically.

Correspondence to: wrcrpp@foxmail.com (Wang Jian Guo)

## 1. Introduction

Unlike polymers with long molecular chains or silica with complex networks of covalent bonds, metals are relatively simple in structure [1]. The bonding and density of metallic melt and crystal are quite similar [2]. Once below the melting point, metallic melts are very likely to crystallize, so metals are admittedly poor glass-formers [1]. In fact, metallic glass (MG) was not synthesized until the 1960s [3]. Although several alloys, e.g. $Pd_{40}Ni_{10}Cu_{30}P_{20}$ [4], $Zr_{41.2}Ti_{13.8}Cu_{12.5}Ni_{10.0}Be_{22.5}$ (Vit1) [5] and $Zr_{46}Cu_{30.14}Ag_{8.36}Al_{8}Be_{7.5}$ [6], can be prepared as bulk MG in size of tens of millimeters at a cooling rate as slow as 1 K/s, most MGs can only be fabricated as ribbons with a thickness of ~ 50 μm at a cooling rate around $10^6$ K/s [7]. So fast cooling rate is hard to be met in engineering practice. As a result, to explore MGs with good GFA remains a momentous challenge in the community.

On the other hand, understanding the glass formation is difficult in physics. It cannot be simply described by the thermodynamic quantities, since the glass formation in metals is a highly nonequilibrium process and the experimental synthesis is strongly affected by external factors in quenching method. However, in the long-term exploration by trial-and-error, a few parameters associated with the glass transition temperature $T_g$, e.g. the reduced glass transition temperature $T_{rg} = \frac{T_g}{T_m}$ ($T_m$ is the melting temperature), the supercooled liquid (SCL) region $\Delta T_x = T_x - T_g$ ($T_x$ is the onset temperature of crystallization), $\gamma = \frac{T_x}{T_g + T_m}$ and $\gamma_m = \frac{2T_x - T_g}{T_m}$ [8], were proposed to identify the GFA in MGs. Unfortunately, exceptions against these parameters are always available [9, 10]. For example, $Cu_{60}Zr_{30}Ti_{10}$ with $T_{rg}$ = 0.62 has a critical thickness, $D_c$, of 4.0 mm, whereas $La_{55}Al_{25}Ni_5Cu_{10}Co_5$ with $T_{rg}$ = 0.57 has a $D_c$ of 9.0 mm [11]. Probably, a single parameter cannot characterize the GFA, $\Psi_g$, in MGs. A full understanding of glass formation may be accessible only in multidimensional space where both the intrinsic properties of material and the applied technique for preparation are taken into account [9].

In fact, Inoue proposed a multidimensional criterion including three empirical items [12], i. e. multicomponent alloy of no less than three elements, greater than 12% atomic radius mismatch and negative mixing heat between the major components, to explore the good glass-formers. Although any of these items is unable to find out the good glass-former individually, they together are in accord with the GFA of almost all the MGs developed up to now. More importantly, Inoue's criterion doesn't concern $T_g$ or $T_x$ which actually depends upon the applied experimental method, so it is essentially a consideration of thermodynamics. Elemental body centered cubic (BCC) metals have been vitrified [13, 14], and their extremely poor GFA seems in favor of Inoue's criterion and Greer's "confusion principle" from the reverse side [12, 15]. The monatomic MG unquestionably demonstrates the kinetic aspect of glass formation, as Zallen [1] summarized Turbull's opinion [16]: *Nearly all materials can, if cooled fast enough and far enough, be prepared as glass.* Recently, although great efforts were devoted to the MG development, a full and deep analysis of the thermodynamic properties, the kinetic behavior and the experimental technique is still lacking. Therefore, one has no choice but to resort to the very time-consuming method of trial-and-error in a, more or less, aimless way when trying to develop a new MG, though one may sometimes draw a little inspiration from the empirical rules mentioned above.

In this work, we investigate the roles the thermodynamics and kinetics play in, and the influence the quenching technique has on, the glass formation of metals in a straightforward manner. Relevant characteristic temperatures are considered and discussed, i.e. casting temperature ($T_0$), melting temperature ($T_m$) at equilibrium, onset temperature of crystallization ($T_x$), glass transition temperature ($T_g$), and the copper-mold temperature ($T_{cm}$) at which the sample is prepared, and $\Delta T_{rg}$ based on $T_{rg}$ is proposed for a rough identification of GFA in glasses. As an influence factors for the cooling rate, thermal diffusivity is also taken into account. Meanwhile, the structural complexity of corresponding crystalline phases in a target composition is closely associated with the likelihood of glass formation. Compared with the relaxation time closely related to the viscosity, the diffusion time is a more important factor for the crystallization. The latter is basically the time window for the glass formation and depends upon the diffusion coefficient and the effective diffusion distance rather than the atomic radius which is closely related to the former. Eventually, a fundamental law that is not recognized yet is supposed to govern the glass formation in physics.

## 2. The cooling process in MG preparation

As a very common method for the preparation of MG, the copper-mold casting is illustrated in Fig.1. The top part of the copper-mold is factually playing the role of a crucible where the metals are melted. After the melt enters the cavity, it will be quenched by the water-cooled copper-mold usually working around room temperature ($T_R$). The cooling process mainly depends upon the heat diffusion. As a function of time $t$ and position $r$, the evolution of temperature $T$ can be described by the thermal diffusion equation

$$\frac{\partial T}{\partial t} = \kappa \nabla^2 T \tag{1}$$

where $\kappa$ is the thermal diffusivity. Consider a rod-shaped sample (see Fig.1) that is axially symmetric and can be idealized infinitely long. For simplification, let's presume the copper-mold keeps the constant temperature, $T_{\rm cm}$. Of course, $T_{\rm cm}$ must vary with time during the quenching process in that the copper-mold absorbs heat from the sample and releases heat to the water, which will be discussed in the following text. To form a fully glassy rod sample with a radius of $R$, one must ensure that the material at the axis ($r$ = 0) is quenched at a cooling rate fast enough because the axis-center material is certainly quenched at the slowest cooling rate. As a consequence, we just focus the region at $r$ = 0 hereafter. If the initial temperature of the melt in the cavity is $T_0$, the temperature $T$ and cooling rate $\dot{T}$ at $r$ = 0 and time $t$ are [17]

$$T = T_{\rm cm} + \frac{2\Delta T}{R} \sum_{n=1}^{\infty} e^{-\kappa \alpha_n^2 t} \frac{J_0(0)}{\alpha_n J_1(x_n)} \tag{2}$$

$$\dot{T} = -\frac{\partial T}{\partial t} = \frac{2\kappa \Delta T}{R^2} \sum_{n=1}^{\infty} e^{-\kappa \alpha_n^2 t} \frac{x_n J_0(0)}{J_1(x_n)} \tag{3}$$

where $J_0$ and $J_1$ are the first kind Bessel function of order zero and one, $x_n$ is the $n$th positive root of $J_0(x) = 0$, $\alpha_n = \frac{x_n}{R}$ and $\Delta T = T_0 - T_{\rm cm}$. Clearly, $\dot{T}$ is in direct linear proportion to $\Delta T$. Although the $\kappa$ and $R$ dependence of $\dot{T}$ is not so straightforward, $\dot{T}$ must be proportional to $\kappa$ and inversely proportional to $R$.

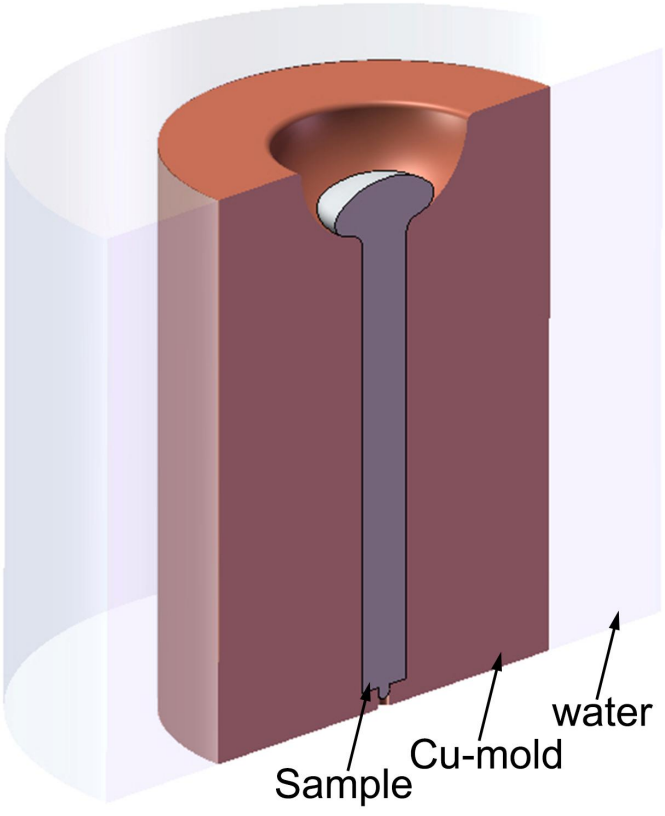

Fig. 1. Illustration for the MG sample preparation by copper-mold casting

The Vit1 ($Zr_{41.2}Ti_{13.8}Cu_{12.5}Ni_{10.0}Be_{22.5}$) has extensively been used as a model material to investigate the physical and mechanical properties of MG due to its excellent GFA and low cost [5]. Here, we also employ Vit1 to study the factors affecting the cooling process of the sample under preparation. MGs and metallic SCLs usually have a thermal diffusivity $\kappa$ of ~$10^{-6}$ $m^2/s$ [18-20]. For Vit1, $\kappa$ ranges from $2.0\times10^{-6}$ $m^2/s$ at 300 K in glass state to $6.0\times10^{-6}$ $m^2/s$ at 700 K in crystalline phase, and $T_{\rm m}$ is 969 K while $T_{\rm g}$ is 623 K [20]. To ensure the fluidity of the melt, the casting temperature $T_0$ is usually at least tens of kelvins higher than $T_{\rm m}$.

First of all, let's investigate the size effect of the sample on the cooling rate $\dot{T}$ with $\kappa$=3.0 $\times10^{-6}$ $m^2/s$ and $T_0$ =1000 K (31 K higher than $T_{\rm m}$). One must note that the temperature zone of $\Delta T_{\rm m} = T_{\rm m} - T_{\rm g}$ is dangerous because the crystallization is likely to happen anytime. In order to form a glass, the melt should go through this dangerous zone as soon as possible. If the elapsed time in this zone is defined as $\tau_g$, the average cooling rate during the period of $\tau_{\rm g}$ is $\dot{T}_{\rm a} = \frac{\Delta T_{\rm m}}{\tau_{\rm g}}$ and therefore $\tau_{\rm g} = \frac{\Delta T_{\rm m}}{\dot{T}_{\rm a}}$. Apparently, a short $\tau_{\rm g}$ is highly desired, which requires a small $\Delta T_{\rm m}$ and a large $\dot{T}_{\rm a}$. Fig.2a shows the evolution of temperature at $r = 0$ in the samples with $R$ of 1, 3, 5 and 10 mm, respectively. As expected, $\tau_{\rm g}$ is smaller for a thin sample than for a thick one. For instance, $\tau_{\rm g}$ for the sample of $R = 1$ mm is $\tau_{\rm g1} = 0.049$ s and for the sample of $R = 10$ mm is $\tau_{\rm g10} = 4.815$ s. The former has an average cooling rate about $10^2$ times faster than the latter, which roughly indicates $\dot{T}_{\rm a} \propto \frac{1}{R^2}$. Lin and Johnson [21] proposed a semi-quantitative relationship between the average cooling rate $\dot{T}_{\rm a}$ and the diameter of rod sample, $2R$, in centimeter as follows

$$\dot{T}_{\rm a} = \frac{C}{R^2} \tag{4}$$

where $C = 40$ was suggested in ref. [21]. Fig.2b shows this relationship, and $C = 72$ in Vit1 is obtained by fitting the data. Of course, $C$ should not be a constant, and it depends upon $\kappa$ and $\Delta T$ [21], as reflected in Eq. (3). Anyway, the reduction of sample thickness can accelerate the cooling rate remarkably and therefore greatly raise the likelihood of the glass formation, consistent with the experiments. On the other hand, a higher $T_g$ will lead to a smaller $\Delta T_m$ if $T_m$ is fixed, which means the dangerous temperature zone shrinks [16]. In Fig.2a, if the $T_g$ is elevated by 100 K (i.e. $T_g$=723 K), $\tau_{g1}$ will decrease from 0.049 s to 0.032 s. Obviously, a higher $T_g$ makes the time window for crystallization squeezed and the access to glass therefore widened.

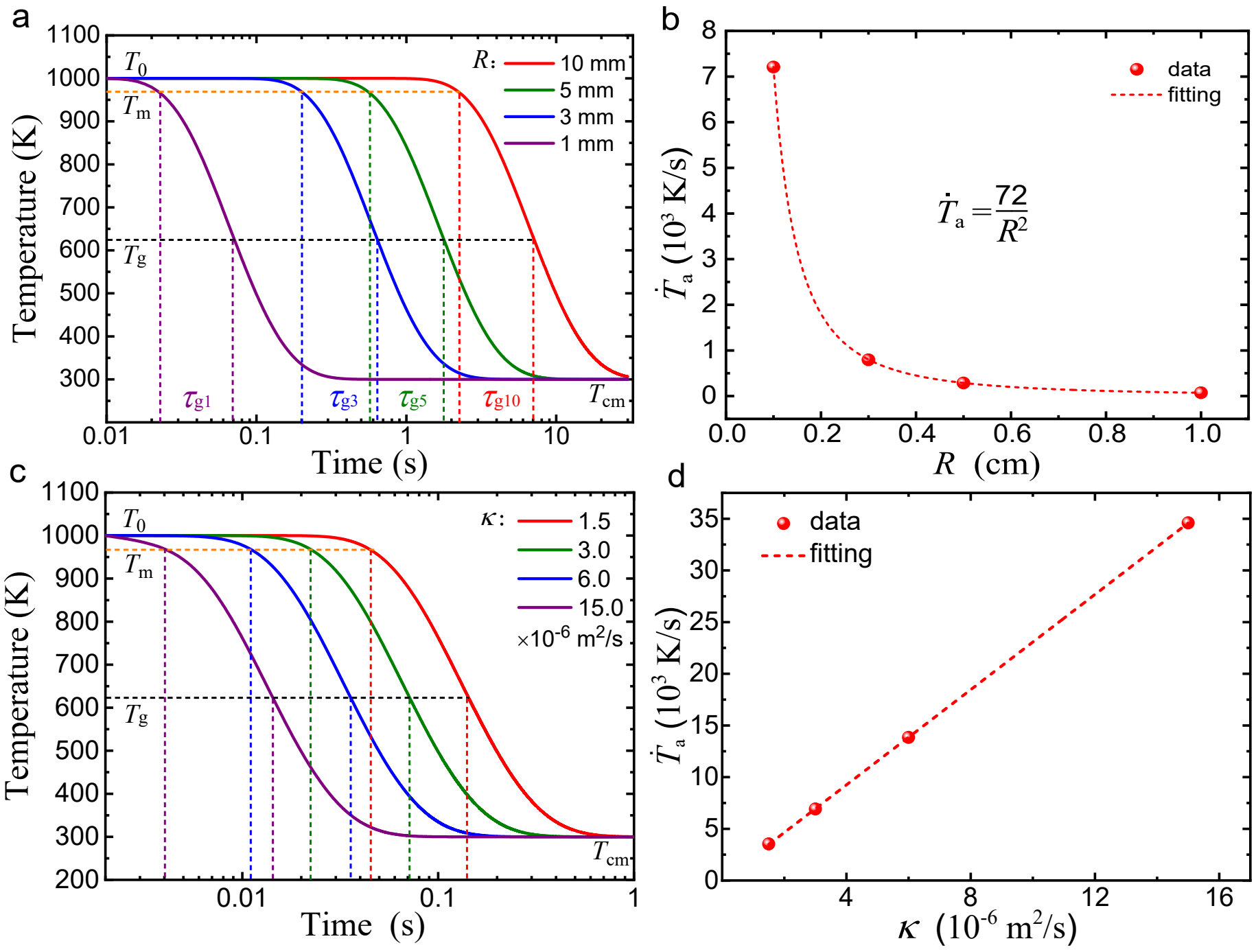

Fig. 2. The influence of sample size $R$ and thermal diffusivity $\kappa$ on the average cooling rate $\dot{T}_a$ in Vit1. (a) $\tau_g$s for $R$ = 1, 3, 5, 10 mm and $\kappa$ = 3.0×10$^{-6}$ m$^2$/s by which the $\dot{T}_a$ is correlated with 1/$R^2$ in (b). (c) $\tau_g$s for $\kappa$ = 1.5, 3.0, 6.0, 15.0×10$^{-6}$ m$^2$/s and $R$ = 1 mm which show a $\kappa$ linear dependence of $\dot{T}_a$ in (d).

According to Eq. (3), a large thermal diffusivity $\kappa$ must help quench the material at a fast cooling rate. Fig. 2c shows the temperature evolution in the sample of $R$ = 1 mm with $\kappa$ = 1.5, 3.0, 6.0 and 15.0×10$^{-6}$ m$^2$/s, respectively. One can easily find that $\tau_g$ decreases with the increasing $\kappa$. The average cooling rate $\dot{T}_a$ during the period of $\tau_g$ is presented in Fig.2d, which corroborates that a large $\kappa$ increases $\dot{T}_a$. On the other hand, the linear relationship between $\dot{T}_a$ and $\kappa$ indicates that increasing $\kappa$ is not so effective as reducing the thickness $R$ to improve the $\dot{T}_a$ (see Fig.2b). At the same time, one should learn that the $\kappa$s for different MGs are almost the same [18], though they differ significantly for various elemental metals [22], which is presented in Fig. S1. As a result, it seems unlikely to improve $\dot{T}_a$ substantially by $\kappa$ in practice.

Since the melt will never solidify above $T_m$, a high $T_0$ (> $T_m$) seems unnecessary for the glass formation as long as the fluidity of the melt is ensured. However, a high $T_0$ will enlarge $\Delta T$ in Eq. (3) and therefore increase $\dot{T}_a$. As presented in Fig.3a, when $T_0$ is elevated from 1000 K to 1100 K and 1300 K, $\tau_g$ is shortened from 0.049 s to 0.044 s and 0.042 s, respectively. This confirms the positive effect of a high $T_0$ on $\dot{T}_a$, provided that $T_{cm}$ is kept all the time. However, if $T_0$ goes up to 2000 K, $\tau_g$ is also about 0.042 s (see the red lines in Fig.3a), almost the same as that for $T_0$ = 1300 K. This indicates that a too high $T_0$ is needless. In addition, overheating, i.e. a high $T_0$ above $T_m$, can dissolve the heterogeneities for nucleation, expand the undercooling and lengthen the crystallization time [23], which will also enhance the glass formation. Of course, there exists a threshold value, $T_t$, for $T_0$, over which the enhancement effect on the glass formation is not strengthened any more and even is weakened instead, because the melt is oxidized much more heavily at very high temperature [23-25]. As a result, $T_t$ should be the upper bound for $T_0$.

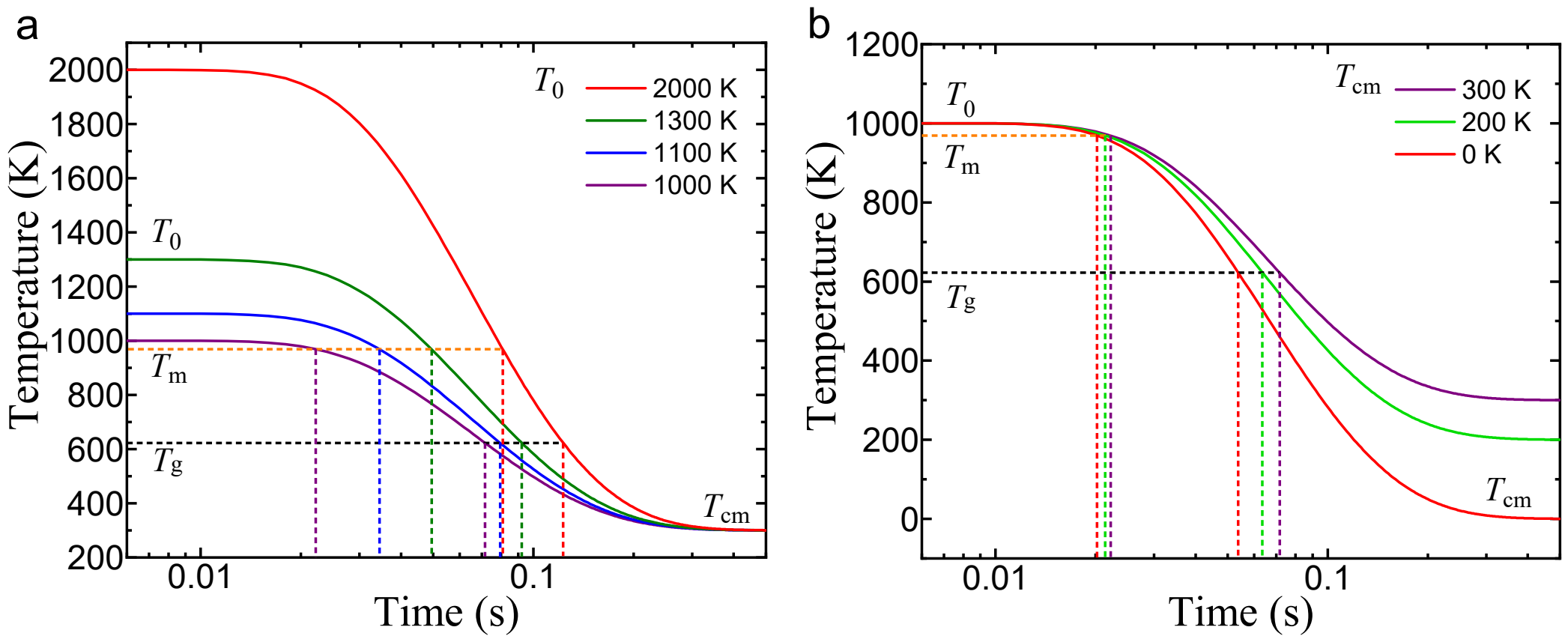

Fig. 3. The influence of casting temperature $T_0$ (a) and Cu-mold temperature $T_{cm}$ (b) on $\tau_g$ which is marked as the width between two vertical dashed lines in the same color.

A lower $T_{cm}$ also leads to a large $\Delta T$, so $T_{cm}$, by intuition, should be applied as low as possible. As shown in Fig.3b, when $T_{cm}$ decreases from 300 K to 200 K and 0 K, $\tau_g$ reduces from 0.049 s to 0.042 s and 0.034 s, respectively. Compared with the elevation of $T_0$, the drop of $T_{cm}$ is more efficient to shorten $\tau_g$. The only pity is that $T_{cm}$ is unable to fall below 0 K, otherwise all matters can be vitrified easily according to Zallen's statement [1]. In addition, $T_{cm}$ actually cannot keep a constant. It, more or less, rises in the cooling process and therefore reduces $\Delta T$ due to the accumulated heat released from the sample. A high $T_0$ will result in a large increment of $T_{cm}$, which also indicates that an applied casting temperature in excess of $T_t$ is unnecessary. On the other hand, a very low $T_{cm}$ may not always be helpful for the glass formation, given the mold-filling capability of the melt in die-casting. If the Cu-mold is very cold, the melt cannot enter the cavity smoothly because it will immediately solidify once touching the inner wall of Cu-mold and the following melt is subject to turbulence [9], which is very likely to cause the cold lap on the surface of the sample. The cold lap separates the sample and the Cu-mold, across whose interface the heat transfer is strongly blocked. Therefore, the cooling rate dramatically decreases, and the risk of crystallization soars. In this sense, the balance or optimality between the temperature gradient (i.e. $T_0$ and $T_{cm}$) and the interfacial heat transfer should be reached to achieve the best cooling effect.

**3. The time for nucleation**

In general, it is believed that vitrification competes with crystallization during the quenching process. In the SCL region, the materials always prefer to crystallize if the duration is long enough, because the crystalline phase has a lower free energy and is thermodynamically more stable. Therefore, the material must go through the SCL region soon enough in order to avoid the crystallization and reach the glass state, i.e. a small $\tau_g$. However, $\tau_g$ cannot be infinitesimal, for the quenching must take some time, as analyzed above. Consequently, some pure metals (e.g. Cu, Au and Pd) fail to form glass even through the state-of-the-art quenching technique [14]. Fortunately, the time needed for crystallization, $\tau_c$, is not infinitesimal either. The sufficient condition for the glass formation is

$$\tau_g < \tau_c \tag{5}$$

Logically, there are two independent ways to make inequality (5) hold. One is to decrease $\tau_g$, which has been analyzed above; the other one is to increase $\tau_c$. Exactly speaking, there should be no any crystallite in a fully glassy material. In this case, $\tau_c$ mainly depends upon the nucleation process, and it is closely related to the characteristic diffusing distance of atoms, $l$, and the diffusion coefficient, $D$ [26]

$$\tau_c \propto \frac{l^2}{D} \tag{6}$$

The $l$ has the same magnitude order as the size of crystallographic unit cell. The unit cell of pure metal is simple and small, typically BCC, FCC or HCP. The metallic compound usually has complex unit cell and large $l$. For instance, pure Cu is of FCC with $l \sim 3.6$ Å and the $Cu_{10}Zr_7$ is a compound with $l \sim 10$ Å. Apparently, multicomponent alloys are easier than elemental metals to form MGs. Turnbull [16] even proposed that the ideal glass has an infinitely big unit cell, i.e. $l \rightarrow \infty$.

In reality, the small crystalline nuclei scattered in a material can hardly be observed in the experiment. They must grow over a critical size for the clear identification of crystallization. Therefore, the growth of nucleus should be taken into consideration when the glass formation is analyzed. In a multi-component alloy, there are usually a number of crystalline phases, so the growth of nuclei concerns the long-distance diffusion of atoms since the compositions for various phases are usually different. Then $l$ in the proportional relation (6) is about the critical size of a grain instead of a unit cell. In addition, the competition among these phases can frustrate the crystallization and then help form glass [27]. In other words, the more the phases are, the better the GFA is. When the effect of pressure is neglected, the number of elements (or compounds), $N$, and the number of coexisting phases, $P$, obey the Gibbs phase rule:

$$P = N - F + 1 \tag{7}$$

where $F$ is the number of degrees of freedom. The maximum number of crystalline phases equals $N$ at the eutectic or peritectic point and $N + 1$ at the eutectoid point when $F = 0$ is taken. As such, for a multi-component or multi-element system, we propose the crystallization time $\tau_c$ is estimated by:

$$\tau_{\mathrm{c}} \propto \sum_{i=1}^{P} f_i \sum_{j=1}^{J_i} C_{M_i}^{m_{ij}} \frac{l_i^2}{D_j} \tag{8}$$

in which $f_i$ (⩽1) is the weight factor of each crystalline phase and $f = 1$ if $P = 1$, $J_i$ is the number of elements in $i$th phase, $m_{ij}$ is the number of atoms of $j$th element in a critical-sized grain of $i$th phase, $M_i = \sum_{j=1}^{J_i} m_{ij}$ is the total number of all atoms belonging to a critical-sized grain of $i$th phase, $C_{M_i}^{m_{ij}} = \frac{M_i!}{m_{ij}!(M_i - m_{ij})!}$ is a rough consideration of atom permutation, $l_i$ is the critical size of grain in $i$th phase, and $D_j$ is the atomic diffusivity of $j$th element in SCL. At the same time, the driving force for crystallization must be taken into account, i.e. the free energy difference between SCL and crystalline solid with the same chemical composition, $\Delta G$. As a result, the expression (8) should be modified as:

$$\tau_c = \tau_0 \sum_{i=1}^{P} f_i \sum_{j=1}^{J_i} C_{M_i}^{m_{ij}} \frac{l_i^2}{D_j} e^{-\frac{\Delta G_i}{k_B T}} \quad (9)$$

in which $\tau_0$ is a prefactor and $k_B$ is the Boltzmann constant. Given the energy barrier to develop the interface between the SCL and the solid, there exists [2]

$$l_i \approx \frac{4\sigma}{\Delta G_i} \quad (10)$$

where $\sigma$ is the interfacial tension. A small $\Delta G$ not only demonstrates a weak driving force for crystallization but also increases the diffusing distance of atoms, both of which can prolong $\tau_c$ in Eq. (9) and augment the GFA. In general, the more elements an alloy consists of, the smaller its $\Delta G$ is. For example, at the temperature of $0.8T_m$, $\Delta G$ for the quinary Vit1 as a robust glass former with a critical cooling rate as slow as 1 K/s is ~1.5 kJ/mol, significantly smaller than ~1.9 kJ/mol for the binary $Zr_{62}Ni_{38}$ as a poor glass former with a critical cooling rate of $10^4$ K/s [28]. One can easily draw a conclusion that the multi-element, as suggested by Inoue [12], is a feature of a good glass-former indeed [15].

Another way to prolong $\tau_c$ in Eq. (8) is to decrease the diffusivity $D$ of atoms. Although the Stokes-Einstein relation is widely found to break down in the regime of SCL [29, 30], one can still be acquainted with the main governing factors of $D$ by

$$D = \frac{k_B T}{6\pi \eta r_{se}} \quad (11)$$

where $\eta$ is the viscosity and $r_{se}$ is the Stokes-Einstein radius which can be estimated by atomic radius. Accordingly, low temperature, high viscosity, and elements with large atomic radius are expected to decrease $D$. To achieve a relatively stable SCL at low temperature, the alloy should have a low $T_m$, since the SCL exists below $T_m$. Eutectic composition normally has a low $T_m$ at which the melt becomes more viscous, i. e. large $\eta$ in Eq. (11) according to the Arrhenius equation, so the eutectic composition should be preferred over other compositions to prepare MG in an alloy. The criterion of eutectic composition was pointed out and emphasized by Cohen and Turnbull [16, 31]. As a solid evidence, a number of binary alloys, e.g. Pd-Si [32, 33], Ni-Nb [34, 35], Ca-Al [36] and Cu-Zr [37, 38], are successfully prepared as bulk MGs around eutectic compositions. Although the off-eutectic compositions in some alloys are found to have better GFA than the eutectic ones, the composition difference between them is not so significant [39-42], so the eutectic criterion is a vital directional guide for designing MGs. On the other hand, the pressure dependence of diffusivity, i.e. *p v.s. D*, has also been studied in SCLs and glasses [43, 44]. Although this dependence is hardly detectable in some MGs, it is usually perceptible in metallic SCLs [45]. It is found that the viscosity increases, and therefore the diffusivity decreases with increasing pressure according to Eq. (11). Schmelzer *et al.* [46] theoretically analyzed the pressure dependence of viscosity in liquids, and they presented that a positive thermal expansion would lead to a positive pressure dependence of viscosity as long as the free-volume model can describe the atomic structure. Although the negative thermal expansion coefficient was suggested in a Ce-based MG under 0.6 GPa according to the shift of first peak position on the intensity curve of scattering by synchrotron [47], the direct measurements in most MGs and SCLs usually show positive thermal expansion [7]. Also, molecular dynamics simulations have clearly shown that the atomic diffusion in Lennard-Jones and metallic SCLs is slowed down by the applied pressure [44, 48]. As a rough estimation, $D \propto e^{-\lambda p}$ is proposed. Jiang *et al.* [49] found that the crystallization is indeed retarded in a Fe-based multi-element MG due to the reduced mobility of atoms by external pressure. In other words, the pressure would facilitate the vitrification. Moreover, the isotope-effect measurement indicates that the atomic mass, $m$, affects the diffusivity as well, which can be described by $D \propto 1/\sqrt{m}$ [45]. This seems to agree with people's intuition: heavy atoms naturally move slowly. The excellent GFA in Zr-based and Pd-based MGs exemplifies the effect of atomic mass [6, 50], since Zr and Pd have relatively large atomic mass. Interestingly, the self-diffusivity of atoms or molecules in gas, $D_g$, also depends upon these factors in a similar way, which is formulated as $D_g = \frac{3(k_B T)^{3/2}}{8\sqrt{\pi} p d^2 \sqrt{m}}$ and $d$ is the atom or molecular diameter [51, 52].

Once the crystallization time, $\tau_c$, is concerned, one would like to refer to the time-temperature-transformation (TTT) curve [53]. The time to crystallize a tiny volume fraction, $x$, at a temperature below $T_m$ is [28]

$$t_x = \left( \frac{3x}{\pi I_s u^3} \right)^{1/4} \tag{12}$$

in which $x$ is often taken to be ~$10^{-6}$ [54], $I_s$ is the steady-state nucleation rate and $u$ is the growth rate

$$I_\mathrm{s} = \frac{I_0}{\eta} \exp\left[ \frac{-16\pi}{3} \frac{\beta^3 \Delta S_\mathrm{m} T^2}{R_\mathrm{g}(T_\mathrm{m} - T)^2} \right] \tag{13}$$

$$u = \frac{u_0 f}{\eta} \left[ 1 - \exp\left( -\frac{\Delta S_\mathrm{m}(T_\mathrm{m} - T)}{R_\mathrm{g} T} \right) \right] \tag{14}$$

$I_0$ and $u_0$ are prefactors, $\beta$ is a factor for atomic arrangement at the interface, $\Delta S_\mathrm{m}$ is the entropy change for melting, $f$ is the fraction of nucleation sites at the interface, and $R_\mathrm{g}$ is the gas constant. If the volume fraction corresponds to the aforementioned critical-size grain, Eq. (12) is equivalent to Eq. (9), i.e. $\tau_\mathrm{c} \Leftrightarrow t_\mathrm{x}$. However, the influencing factors for $I_s$ and $u$ usually need to be measured experimentally. Therefore, Eq. (12) can hardly be taken advantage of to design the composition of MG. Eq. (9) considers the crystallographic features, so it provides some clues for the exploration of possible MGs. Owing to the metastable nature of the SCL, the crystallization events occur in a random manner [53]. It is reasonable to presume that the crystallization time or frequency follows a stochastic distribution, such as Gaussian distribution. Either Eq. (9) or Eq. (12) actually presents the mathematical expectation of the crystallization time. Fig. 4a schematically shows the possibility density, $f(t, T)$, of transformation with time and temperature below $T_\mathrm{m}$. This 3-dimensional TTT diagram indicates that the SCL will crystallize at any time just with different probabilities. Fig. 4b is the projection of $f(t, T)$ in the $t$-$T$ coordinate plane, which actually is like an extended TTT curve. The overall risk of crystallization can be accessed by the cumulative distribution function

$$F_s = \int_s f(t,T)ds = \int_0^t f(t,T)\sqrt{1 + (\frac{\partial T}{\partial t})^2}\, dt \tag{15}$$

where $s$ marks the cooling path, and two paths are labeled in Fig. 4b. Since the integrand, although it may be very small in absolute value, is positive in Eq. (15), $F_s$ must increase with time. The crystallinity in a glassy material, as suggested by Uhlmann [54], is generally less than $10^{-6}$, so $F_s$<$10^{-6}$ must be satisfied when the temperature falls below $T_g$ in the cooling process. On the other hand, $F_s$ depends on the cooling path. For example, $F_s$ for path 1 is greater than that for path 2 in Fig. 4b even though they share the same onset and destination, so path 2 should be preferred to reduce the crystallinity. In turn, the fully glassy (i.e. crystallinity-free) state in MGs seems not available in principle unless the ground state of an alloy is of glass. This leads to the poor reproducibility of MGs to some extent, which is reflected by the scattered data of strength, plasticity and relaxation enthalpy in MG [55].

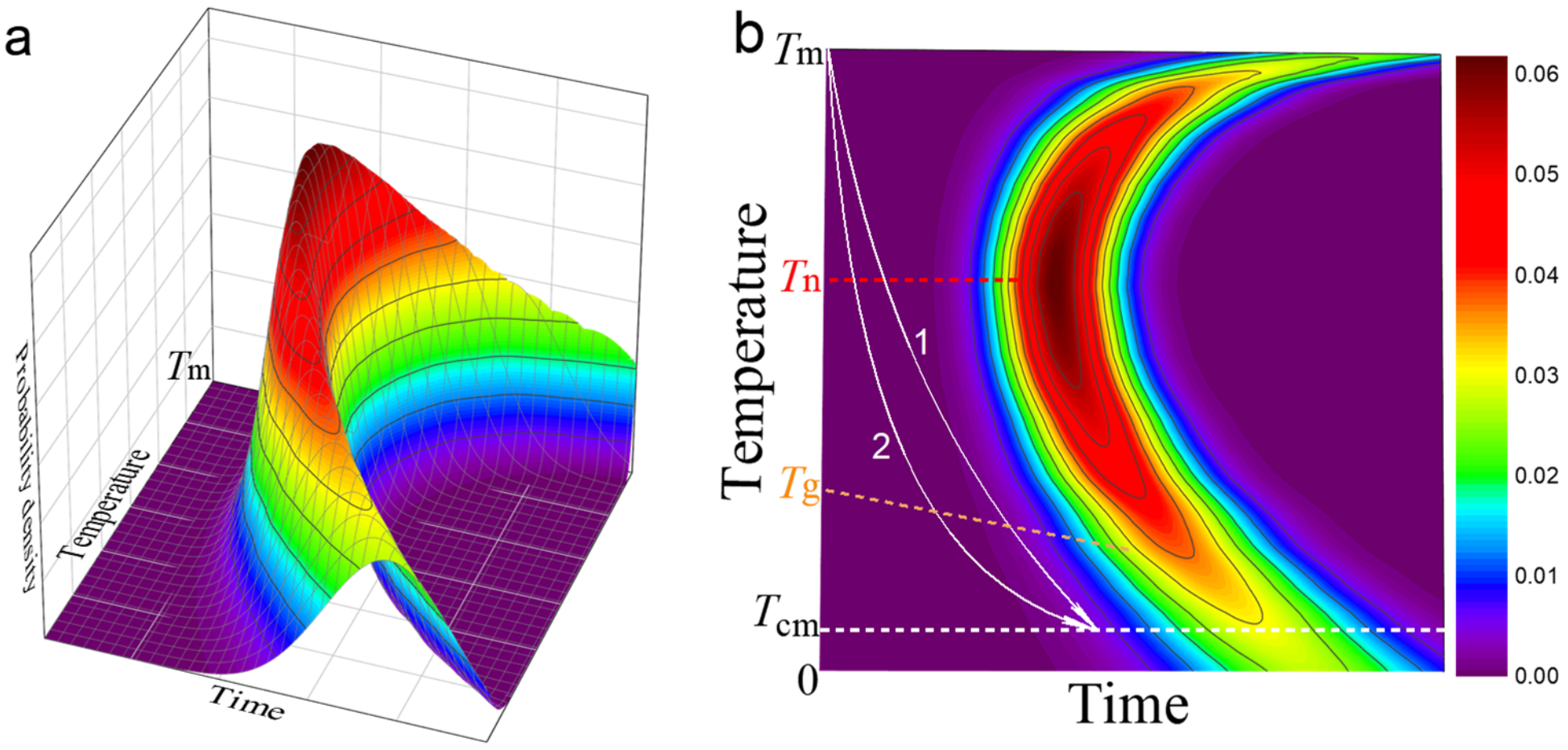

Fig. 4. An illustration for TTT-diagram. (a) The 3D TTT-diagram with a ridge representing the nose-shaped curve for crystallization time in conventional TTT-diagram. (b) The projection of (a) in *t*-*T* plane. The color scale represents the probability density for crystallization. $T_n$ is the nose temperature.

## 4. Consideration of phase diagram

Phase diagram is by far the most useful guidance for selecting the composition of MG [56]. The preceding eutectic criterion is of course based on the phase diagram. Unfortunately, only the elemental phase diagrams have been studied completely [57], and phase diagrams for multi-element alloys are not always available or accurate. The binary alloy Ca-Al, for instance, has recently been found to have two new complex compounds $Ca_{14}Al_{13}$ and $Ca_8Al_3$ [36], between which there exists a eutectic point $Ca_{66.2}Al_{33.8}$, and a number of bulk MGs are synthesized near this point. Although the phase diagram provides the phase-transition temperature (e.g. $T_m$) and crystallographic data which help to estimate $\tau_g$ and $\tau_c$ through Eqs. (2), (3) and (9), its usage, particularly the eutectic criterion as aforementioned, is contentious [16, 31, 42, 58, 59]. Admittedly, there is no any bulk MG fabricated in a binary alloy system which doesn't show any eutectic point in its phase diagram [60-62]. This disproof corroborates the implication of the eutectic criterion and therefore the phase diagram for designing MGs. Therefore, the phase diagram, particularly with the eutectic, is worth more attentions.

Fig. 5 presents the schematic phase diagram of a binary A-B alloy system with three eutectic points. The eutectic temperatures are $T_1$, $T_2$ and $T_3$, respectively. At first glance, one may pick the composition of AB, compared with $C_1$ and $C_3$, as a better glass-former according to the empirical rule of deep eutectic. From the perspective of equilibrium thermodynamics, the melt AB will solidify to two compounds of $A_mB_n$ ($m > n$) and $A_pB_q$ ($p < q$) at the temperature of $T_2$. Obviously, $A_mB_n$ or $A_pB_q$ must have a more complex and bigger unit cell than elemental A or B. $\tau_c$ for AB is therefore longer than that for $C_1$ or $C_3$ according to Eqs. (9) and (10). On the other hand, if $A_mB_n$ has a larger $l$, a smaller $\Delta G$ and more atoms (i.e. a larger $M$) in Eq. (9) than $A_pB_q$, the optimum composition for glass should be shifted slightly away from the eutectic AB toward $A_mB_n$ in order to suppress the primary nucleation of $A_pB_q$ caused by constituent fluctuation, and that's $AB'$. Li *et al*. [40, 59, 63] attributed the superior GFA of off-eutectic glass to the asymmetric eutectic coupled zone where the structure varies with the growth rate and temperature, and they concluded that the best GFA can be fulfilled on the side with a steeper liquidus line. However, Xia *et al*. [42, 64] found that the better glass former is on the side with a gentle liquidus slope in Cu-Hf binary alloy. They calculated the formation enthalpies of glass ($\Delta H_g$) and intermetallic compounds ($\Delta H_c$) and suggested that a large $-\Delta H_g$ and a small $\Delta H_g-\Delta H_c$ would achieve a good GFA. $\Delta H_g-\Delta H_c$ is equivalent to $\Delta G$ in Eq.(8) if the contribution of entropy is negligible [42]. Xia *et al*'s consideration is in accordance with our scenario, but they didn't take into account the size $l$ and the number of atoms $M$ of a critical-size nucleus. In fact, it is difficult to measure $l$ and $M$ of a critical-size nucleus experimentally in that these two parameters depend on the supercooling degree of the melt. As an alternative, the unit cells can be used to roughly compare $l$ and $M$ for different phases in practice. Since the crystallographic data of phases can be referenced according to the phase diagram, $l$ and $M$ for the unit cell will be determined easily. A larger unit cell usually contains more atoms, i.e. $l \propto M$. Therefore, $M$ is a key and convenient parameter that reflects the complexity of a crystalline phase to a great extent. There are $M = Z(m+n)$ atoms in the $A_mB_n$ unit cell and $Z$ is the number of formula units. For example, $Z = 4$ is for the $Cu_{10}Zr_7$ unit cell which then contains 68 atoms while $Z = 1$ is for the $Cu_{51}Zr_{14}$ unit cell which does 65 atoms. A robust glass-former, once crystallized, should produce a number of crystalline phases of large $M$. As shown by the dashed curve in Fig. 4a, if the deep eutectic is located at $C_2$ (close to $A_pB_q$) instead of AB, it is typically the asymmetric eutectic emphasized by Li *et al.* [40]. Now the optimum composition for glass formation is off-eutectic and keeps a little away from the $A_pB_q$ end, e. g. $C'_2$, provided that $M$ for $A_mB_n$ is no less than that for $A_pB_q$. This is due to the small composition difference between $C_2$ and $A_pB_q$ which must facilitate the precipitation of $A_pB_q$. Indeed, Wang *et al.* [65] found that the intermetallic compositions are marginal glass former in binary Cu-Zr alloy system. However, if the complexity (i.e. $M$) for $A_pB_q$ is much greater than that for $A_mB_n$, the GFA of a composition (e.g. $C''_2$) approaching $A_pB_q$ will be enhanced, even up to that of $C'_2$. There may develop a range from $C'_2$ to $C''_2$ where GFA is not so sensitive to the composition any longer due to the combined effect of structural complexity and compositional difference.

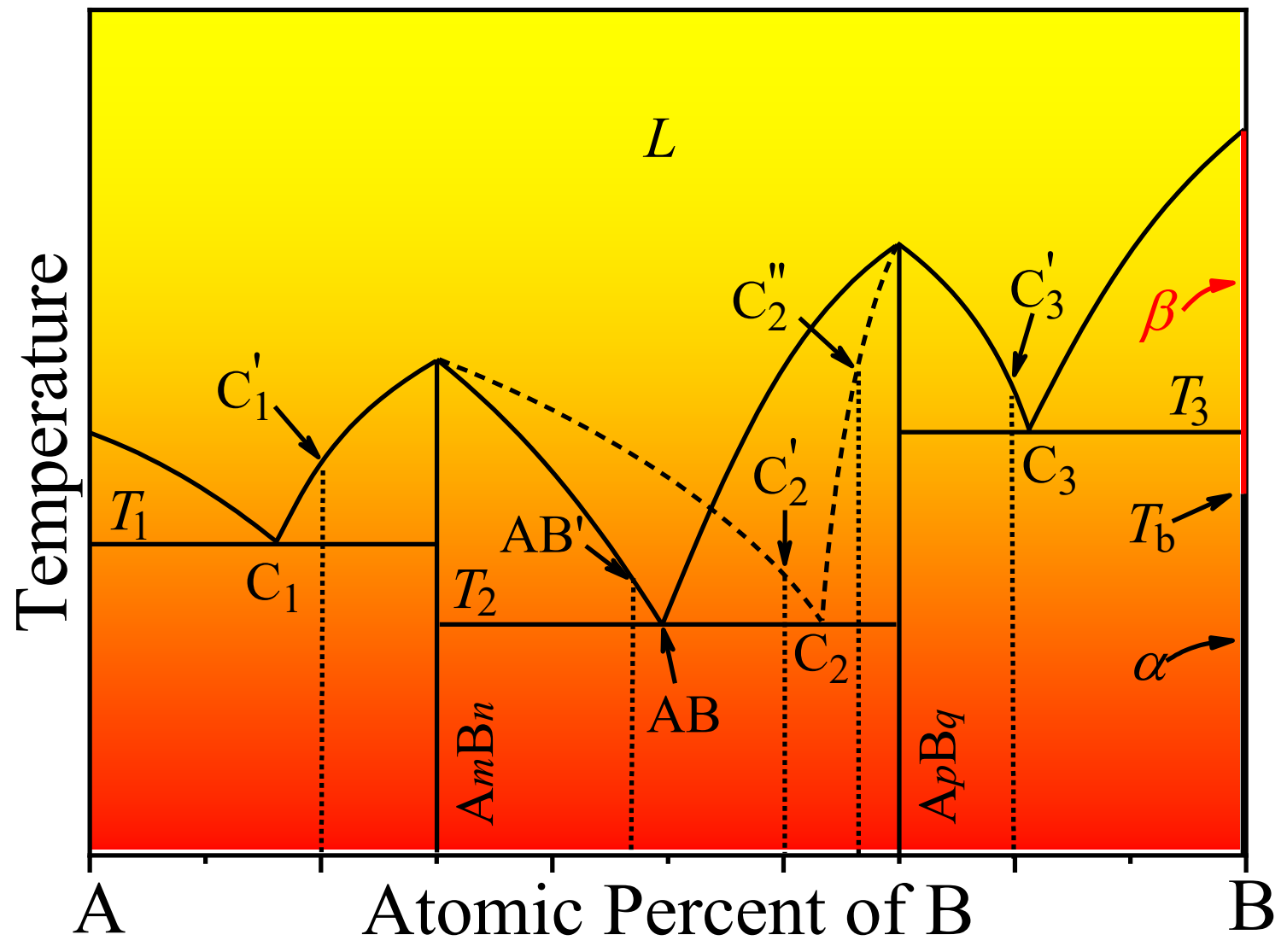

Fig. 5. A typical phase diagram in A-B with eutectic points. Elemental B has two crystalline phases, $\alpha$ and $\beta$, between which the transformation happens at $T_b$.

Compared with AB or $C_2$, the eutectic $C_1(C_3)$ is a poorer glass-former because it has an equilibrium phase of pure A(B) which is very simple in structure with a small $M$. Moreover, B has two crystalline phases ($\alpha$ and $\beta$) and the polymorphic transformation occurs at the temperature of $T_b$. Following the above logic, the best glass former around $C_1(C_3)$ should be shifted, more or less, away from the eutectic itself and toward the $A_mB_n$ ($A_pB_q$) end, e.g. $C'_1$

($C_3'$). On the other hand, $\beta$ phase may exist even below $T_b$ in B as long as the continuous cooling is fast, like the retained austenite in steel [66]. $P$ = 3 ($A_pB_q$, $\alpha$ and $\beta$) in Eq. (9) can be granted for the composition around $C_3$ while $P$ = 2 (A and $A_mB_n$) for that around $C_1$. Therefore, the compositional difference between $C_3'$ and $C_3$ is probably less than that between $C_1'$ and $C_1$. This is consistent with the structural confusion from multiple phases [27]. In binary Fe-B alloy, there is a Fe-rich eutectic $Fe_{83}B_{17}$, but even $Fe_{88}B_{12}$ (5 at.% shift from the eutectic toward Fe) can be prepared as glassy ribbon [67, 68]. The reason might be attributed to the triple polymorphism ($\alpha$, $\gamma$ and $\delta$) in crystalline Fe, i.e. $P$ = 4 in Eq. (9).

**5. The modification of $T_{rg}$ criterion**

The melting temperature, $T_m$, always available in the phase diagram of alloys and often used to establish criteria like $T_{rg}=T_g/T_m$ for characterizing the GFA of MGs, is a key parameter itself. A low $T_m$ can elevate $T_{rg}$ and reduce the diffusivity $D$ in Eq. (11), and improve the GFA. However, $T_m$ is the upper limit of $T_g$, and a low $T_m$ will lead to a lower $T_g$. Once $T_g$ is close to or even below $T_{cm}$ (~ $T_R$, the room temperature), the SCL will likely crystallize rather than vitrify since it is not cooled *far enough*. Indeed, there is no any reported MG with $T_g < T_R$ so far [69], because the Cu-mold usually works around $T_R$. On the other hand, the low $T_m$ (~ $T_0$) also results in a small $\Delta T$ in Eq. (3), which will slow down the cooling rate, and the melt cannot be cooled *fast enough*. Therefore, a very low $T_m$ is not in favor of GFA.

Although $T_m$ is a thermodynamic quantity of a crystalline material, it varies with the pressure [57]. In general, the larger the pressure is applied, the higher the $T_m$ increases. As shown schematically in Fig. 6, $T_m$ increases from $T_{ma}$ to $T_{mb}$ due to the applied pressure. This holds true for most compounds [70] and elemental substances [57] as long as the pressure is not very high, usually less than 10 GPa. When the applied pressure increases over 40 GPa, the pressure dependence of $T_m$ becomes weak and $T_m$ almost keeps a constant in some metals [71]. Similarly, $T_g$ is also a thermodynamic quantity for glass, though it varies with the cooling rate, $\dot{T}$. As shown in Fig. 6, $T_g$ increases from $T_{ga}$ to $T_{gb}$ due to the increased $\dot{T}$. Typically, $\dot{T}$ increased by one order of magnitude will cause an increment of no more than two kelvins for $T_g$ [1]. To form a glass, $\dot{T}$ must be no less than a critical value, $\dot{T}_c$, as emphasized above. The maximum value of $\dot{T}$ is by far on the magnitude order of $10^{14}$ K/s [14]. As a result, the range of $T_g$ is around ten kelvins [72-74]. For an MG prepared at ambient pressure, $T_{rg}=T_g/T_m$ is therefore a quite reliable parameter for roughly describing the GFA [75]. The crystallization temperature $T_x$ is often taken to identify the GFA in MG, e.g. $\Delta T_x=T_x-T_g$ [8]. In fact, $T_x$ must be absent during the process of glass formation, otherwise crystallization must occur and the MG will fail to form. In other words, $T_x$ should not be related to GFA directly, because it has nothing to do with the glass formation which occurs in cooling process, and $T_x$ is usually measured in heating process of MG. It actually measures the thermal stability of an MG. Compared with $T_g$ or $T_m$, $T_x$ has a wider range, and $T_g < T_x < T_m$ is expected. If the heating rate is fast enough, e.g. 200 K/s for Vit1 [76], the crystallization can be avoided, so $T_x$ is undetectable. Obviously, beingness and value of $T_x$ entirely depend upon the heating rate. In this sense, $T_x$, unlike $T_g$, is a completely kinetic parameter for MG, and its adoption in identification of GFA may be not fundamentally underpinned in physics. On the contrary, the temperature of copper mold, $T_{cm}$, has never been taken into account to describe the GFA of an MG [8], since $T_{cm}$ is an external factor in preparation rather than an intrinsic property in MG. However, as aforementioned, $T_g$ must be higher than $T_{cm}$, which may mean that a large $\Delta T_g=T_g-T_{cm}$ is helpful for glass formation. Then, following Turnbull [16], we propose a criterion for GFA in glasses, particularly in MGs

$$\Delta T_{rg}=\frac{\Delta T_g}{\Delta T_m}=\frac{T_g-T_{cm}}{T_m-T_g} \tag{16}$$

In principle, a robust glass-former must have a greater $\Delta T_{rg}$ than a marginal one. For a material, $\Delta T_{rg}\leqslant 0$ must arise from $T_g\leqslant T_{cm}$ since $T_m> T_g$ is always true, which implies that this material will not form a glass unless $T_{cm}$ goes below $T_g$. In sharp contrast, if $\Delta T_m\rightarrow 0$ (i.e. $T_g\rightarrow T_m$), the material has an infinite $\Delta T_{rg}$, i.e. $\Delta T_{rg}\rightarrow\infty$, and the glass made from this material has the supreme GFA and can hardly be crystallized due to the tiny undercooling required. It can be named *ever-glass* as the counterpart of ordinary glass. Interestingly, for $\Delta T$ in Eq. (3), it has $\Delta T=\Delta T_g+\Delta T_m$ if $T_0$ is replaced by its approximation $T_m$. To increase $\Delta T_{rg}$ and $\Delta T$ (therefore $\dot{T}$) simultaneously, it needs to raise $T_g$ and drop $T_{cm}$ since $T_m$ is invariant under a certain pressure. Given the limited range of $T_g$, dropping $T_{cm}$ may be more efficient in practice, consistent with the presented in Fig. 3. One must keep in mind that the lower bound of $T_{cm}$ is

0 K. Once $T_{cm}$ =0 K, Eq. (16) is

$$\Delta T_{rg} = \frac{T_g}{T_m - T_g} = \frac{T_{rg}}{1 - T_{rg}} \tag{17}$$

At this moment, $\Delta T_{rg}$ is determined by $T_{rg}$ only, therefore totally equivalent to $T_{rg}$. In other words, Turnbull's $T_{rg}$ criterion should be based on $T_{cm}$ = 0 K [16]. However, no any MG has been prepared under $T_{cm}$=0 K so far, which weakens the robustness of the $T_{rg}$ criterion. A large $\Delta T_{rg}$ will lead to a small $I_s$ in Eq. (13) and a small $u$ in Eq. (14) at $T = T_g$, so $\Delta T_{rg}$ is a more fundamental index for GFA in physics. This was also recognized before by Chen *et al.* [77]

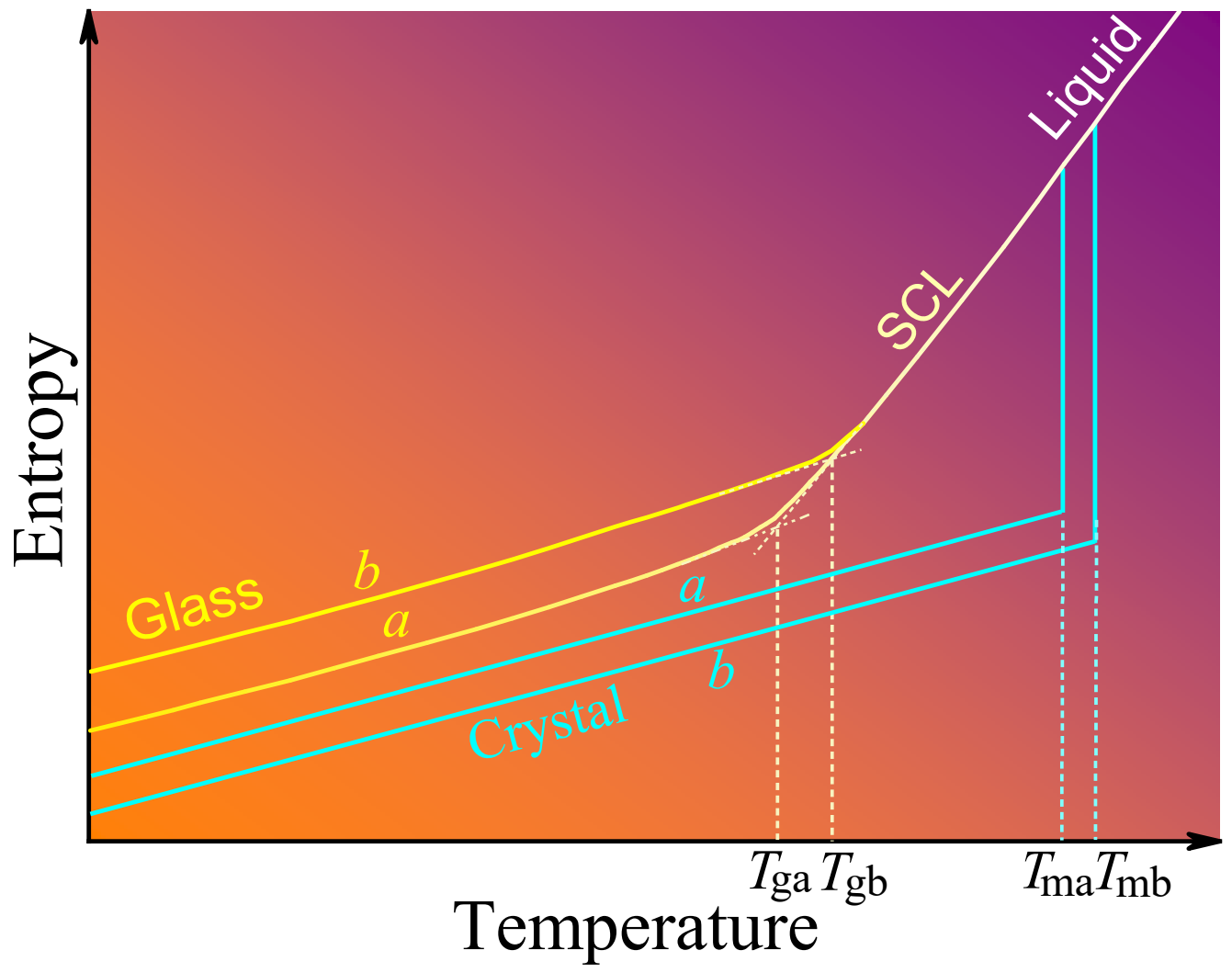

Fig. 6. Temperature dependence of entropy in a liquid. A faster cooling rate will lead to a higher $T_g$ while an applied pressure will elevate $T_m$.

In general, MGs are prepared around room temperature, i.e. $T_{cm} = T_R = 300$ K. In order to check the validity of $\Delta T_{rg}$ criterion proposed in Eq. (16), the $T_g$, $T_m$ and $D_c$ data for a number of representative MGs are listed in Table 1. For comparison, $T_{rg}$ versus $D_c$ and $\Delta T_{rg}$ versus $D_c$ are plotted in Fig. 7 where the MGs are divided into two groups. Some MGs with $T_m < 1000$ K are represented by red circles while others with $T_m > 1000$ K by green ones. At first sight, one can hardly see any clear regularity in Fig.7a or 7b in that the data are so scattered. As discussed above, $T_{cm}$ will rise above 300 K during the melting and then quenching, and a higher $T_m$ leads to a larger rise of $T_{cm}$. To alleviate the influence of $T_{cm}$ rise, only the MGs with $T_m < 1000$ K are selected to correlate $T_{rg}$ and $D_c$ in Fig.7a as well as $\Delta T_{rg}$ and $D_c$ in Fig.7b. In Fig.7a, a linear relationship (dashed line) between $T_{rg}$ and $D_c$ is fitted, and the coefficient of determination (COD), $R^2$= 0.86, is obtained. A few MGs with $T_m > 1000$ K are also around the line, whereas others deviate from the line, especially $Ir_{33}Ni_{28}Ta_{39}$ with $T_m$ =1705 K marked by an arrow. Fig.7b presents the linear fitting for $\Delta T_{rg}$ and $D_c$ with $T_{cm}$ = 300 K, and the COD, $R^2$= 0.94, is obtained. Obviously, $\Delta T_{rg}$ with consideration of $T_{cm}$ has a better linear relationship with $D_c$ than $T_{rg}$. More importantly, most MGs with $T_m >$ 1000 K (green circles) are clearly separated from those with $T_m <$ 1000 K (red circles) around the fitting line. In particular, the $Ir_{33}Ni_{28}Ta_{39}$ with $T_m$ =1705 K goes further away from the line. It demonstrates the necessity of introducing $T_{cm}$ in Eq. (16). However, $T_g$ is usually measured in heating process by differential scanning calorimetry (DSC), rather than cooling process for preparation of MGs. The measured $T_g$ varies more sensitively and perceptibly with heating rate than with cooling rate [85]. In addition, $D_c$ is affected by a few factors mentioned above which can hardly be controlled consistently in different experiments by different investigators using different techniques, so there is no common standard yet on the accurate measurement of $D_c$ in the community so far. Accordingly, the correlation presented in Fig. 7 is not so convincing. More fundamentally, these characteristic temperatures (i.e. $T_g$, $T_m$ and $T_{cm}$) are, perhaps, unable to properly describe the GFA which must be characterized by a number of parameters, and the temperature is only one of them. Anyway, Turnbull's $T_{rg}$ criterion, accompanied by his advice of eutectic composition [16], has shed a light on the rugged way to explore new glasses, and the consideration of $T_{cm}$ helps clean up people's confusion why the light is not so bright sometimes. In mathematics, since $T_{rg} < 1$ is always true, $\Delta T_{rg}$ in Eq. (17) under $T_{cm}$ = 0 K can be expressed with a simple Maclaurin series

$$\Delta T_{rg} = \sum_{k=1}^{\infty} T_{rg}^{k} \tag{18}$$

which indicates that $T_{rg}$ is the first-order approximation of $\Delta T_{rg}$.

Table 1. Summary of data on compositions, properties and critical diameters of MGs.

| Alloy | $T_g$ (K) | $T_m$ (K) | $T_{rg}$ | $\Delta T_{rg}$ | $D_c$ (mm) | Ref. |
|---|---|---|---|---|---|---|
| 1. $La_{66}Al_{14}Cu_{20}$ | 395 | 731 | 0.540 | 0.282 | 2 | [11] |
| 2. $Mg_{61}Cu_{28}Gd_{11}$ | 422 | 737 | 0.572 | 0.387 | 12 | [78] |
| 3. $Nd_{61}Al_{11}Ni_8Co_5Cu_{15}$ | 445 | 744 | 0.598 | 0.484 | 6 | [11] |
| 4. $Mg_{65}Cu_{25}Y_{10}$ | 424.5 | 770.9 | 0.550 | 0.359 | 7 | [11] |
| 5. $Nd_{60}Al_{15}Ni_{10}Cu_{10}Fe_5$ | 430 | 779 | 0.551 | 0.372 | 5 | [11] |
| 6. $Mg_{75}Ni_{15}Nd_{10}$ | 450 | 789.8 | 0.569 | 0.441 | 2.8 | [11] |
| 7. $Mg_{65}Ni_{20}Nd_{15}$ | 459.3 | 804.9 | 0.570 | 0.460 | 3.5 | [11] |
| 8. $La_{55}Al_{25}Ni_5Cu_{10}Co_5$ | 465.2 | 822.5 | 0.565 | 0.462 | 9 | [11] |
| 9. $Pd_{42.5}Cu_{30}Ni_{7.5}P_{20}$ | 575 | 825 | 0.696 | 1.100 | 80 | [50] |
| 10. $La_{55}Al_{25}Ni_{10}Cu_{10}$ | 467.4 | 835 | 0.559 | 0.455 | 5 | [11] |
| 11. $Pd_{40}Cu_{30}Ni_{10}P_{20}$ | 576.9 | 836 | 0.690 | 1.068 | 72 | [11] |
| 12. $Mg_{70}Ni_{15}Nd_{15}$ | 467.1 | 844.3 | 0.553 | 0.443 | 1.5 | [11] |
| 13. $Mg_{80}Ni_{10}Nd_{10}$ | 454.2 | 878 | 0.517 | 0.363 | 0.6 | [11] |
| 14. $La_{55}Al_{25}Cu_{20}$ | 455.9 | 896.1 | 0.508 | 0.354 | 3 | [11] |
| 15. $La_{55}Al_{25}Ni_{20}$ | 490.8 | 941.3 | 0.521 | 0.423 | 3 | [11] |
| 16. $Pd_{40}Ni_{40}P_{20}$ | 590 | 991 | 0.595 | 0.723 | 25 | [11] |
| 17. $Zr_{41.2}Ti_{13.8}Cu_{12.5}Ni_{10}Be_{22.5}$ | 623 | 996 | 0.625 | 0.865 | 50 | [11] |
| 18. $Pd_{77.5}Cu_6Si_{16.5}$ | 637 | 1058.1 | 0.602 | 0.800 | 1.5 | [11] |
| 19. $Pd_{79.5}Cu_4Si_{16.5}$ | 635 | 1086 | 0.584 | 0.742 | 0.75 | [11] |
| 20. $Al_{40}La_{35}Y_{10}Ni_{15}$ | 586 | 1092 | 0.536 | 0.565 | 1 | [79] |
| 21. $Pd_{81.5}Cu_2Si_{16.5}$ | 633 | 1097.3 | 0.576 | 0.717 | 2 | [11] |
| 22. $Zr_{46}Cu_{30.14}Ag_{8.36}Al_8Be_{7.5}$ | 705 | 1103 | 0.639 | 1.017 | 73 | [6] |
| 23. $Pd_{77}Cu_6Si_{17}$ | 642.4 | 1128.4 | 0.569 | 0.704 | 2 | [11] |
| 24. $Cu_{54}Zr_{27}Ti_9Be_{10}$ | 720 | 1130 | 0.637 | 1.024 | 5 | [11] |
| 25. $Pd_{73.5}Cu_{10}Si_{16.5}$ | 645 | 1135.9 | 0.567 | 0.702 | 2 | [11] |
| 26. $Zr_{57}Ti_5Al_{10}Cu_{20}Ni_8$ | 676.7 | 1145.2 | 0.590 | 0.804 | 10 | [11] |
| 27. $Cu_{60}Zr_{30}Ti_{10}$ | 713 | 1151 | 0.619 | 0.942 | 4 | [11] |
| 28. $Pd_{71.5}Cu_{12}Si_{16.5}$ | 652 | 1153.6 | 0.565 | 0.701 | 2 | [11] |
| 29. $Ni_{68.6}Cr_{8.7}Nb_3P_{16}B_{3.2}Si_{0.5}$ | 678 | 1157 | 0.586 | 0.789 | 17 | [80] |
| 30. $Zr_{65}Al_{7.5}Cu_{17.5}Ni_{10}$ | 656.5 | 1167.6 | 0.562 | 0.697 | 16 | [11] |
| 31. $Ti_{34}Zr_{11}Cu_{47}Ni_8$ | 698.4 | 1169.2 | 0.597 | 0.846 | 4.5 | [11] |
| 32. $Fe_{80}P_{11}C_9$ | 690 | 1255 | 0.549 | 0.690 | 1.5 | [81] |
| 33. $Ti_{50}Ni_{24}Cu_{20}B_1Si_2Sn_3$ | 726 | 1310 | 0.554 | 0.729 | 1 | [11] |
| 34. $Fe_{48}Cr_{15}Mo_{14}C_{15}B_6Y_2$ | 848 | 1453 | 0.583 | 0.905 | 9 | [82] |
| 35. $Cr_{50}Co_{29}Nb_7B_{14}$ | 860 | 1517 | 0.566 | 0.852 | 1 | [83] |
| 36. $Ir_{33}Ni_{28}Ta_{39}$ | 1156 | 1705 | 0.678 | 1.559 | 3 | [84] |

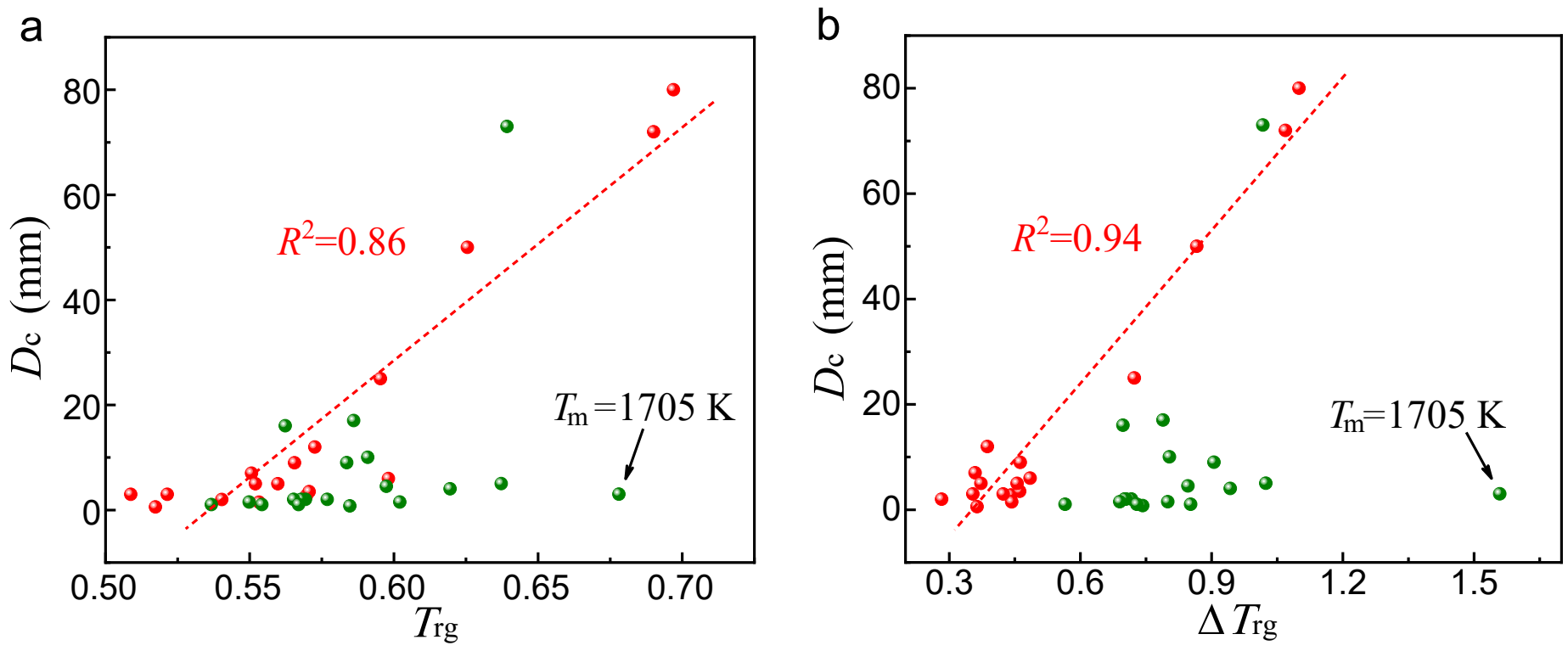

Fig.7. Correlation between critical diameter $D_c$ and $T_{rg}$ (a), and between $D_c$ and $\Delta T_{rg}$ (b). The linear fitting is conducted only using MGs with $T_m$< 1000 K in Table 1.

## 6. Energy landscape and ever glass

Stillinger *et al.* [86-88] proposed the energy landscape (EL) to understand the supercooled liquid and glass as well as glass transition. It presents an intuitive and instructive topographic view for the energy state and dynamics in glass. Following Stillinger, we reanalyze the EL scenario herein. In principle, the EL is of course multidimensional, as emphasized by Stillinger [86-88]. For the convenience of presentation, it is simplistically illustrated in 2-dimension, as shown in Fig. 8. The landscape emerges from the upper-right corner where the liquid phase in thermodynamic equilibrium state is put to an end. When the temperature continues to fall, the melt comes to the SCL which can reach a metastable equilibrium state as long as the perturbation or fluctuation is small. In other words, the equilibrium state in SCL is not stationary. SCL has a supercooling, $\Delta T_e = \frac{\Delta G_e}{k_B}$ ($\Delta G_e$ is the energy difference between the SCL and the liquid at $T_m$). Once $\Delta T_e$ is greater than $\frac{\Delta G_e}{k_B}$, the SCL will vitrify in glass regime if quenched fast or crystallize in crystal regime if cooled slowly in experimental time-scale, as illustrated in Fig. 8. For comparison, Fig.8a presents the EL for a marginal glass-former while Fig.8b does that for a robust one. First of all, $\Delta G_e$ is larger for the robust than for the marginal, which must lead to a larger $\Delta T_e$. Of course, for different alloys with different $T_m$, the reduced supercooling $\Delta T_{er} = \Delta T_e / T_m$ should be compared instead. Secondly, the glass regime in Fig. 8a is narrower than that in Fig. 8b. Some substances, such as pure He [16], cannot vitrify at all, so their glass regime must be almost zero wide. On the contrary, others, like atactic polymer [89], can hardly crystallize, so their glass regime should cover the whole EL and therefore the crystal regime is nearly absent. Because the SCL is structurally similar to glass, it enters the glass regime at first during the cooling. Stillinger suggested that each point on the abscissa (i.e. the configuration) represents a specific atomic arrangement of the whole sample [87]. From the perspective of cluster, different configuration also reflects the atomic arrangement for different cluster. If two clusters share the same configuration, they are equivalent in atomic structure and occupy the same point on the abscissa in Fig. 8. The size of the cluster is on the magnitude order of $l$ in Eq. (9). Apparently, there are countless clusters in a real sample, and all the configurations are occupied by these clusters in EL. Some configurations are occupied by more clusters while others are done by fewer. In other words, the number of clusters occupying every configuration has a distribution. Different samples can be distinguished from each other by the different EL, and the same sample at different energy states, e.g. as-prepared, deformed or annealed, can be distinguished by the varied distribution. There are fewer energy minima, i.e. the configurations like G-I, G-II and G-III, in the glass regime for the marginal glass-former (Fig. 8a) than for the robust one (Fig. 8b). The configuration G-I is very close to the crystal regime, so crystallization will happen to G-I clusters first if the cooling is not fast enough or the glass is annealed. Compared with the EL in Fig. 8b, there are more G-I clusters in Fig. 8a due to the fewer configurations in the narrower glass regime. In addition, the driving force for crystallization or energy difference between glass and its crystal is greater in the marginal glass than in the robust glass. Therefore, it is easier for these clusters to agglomerate together and even percolate through the whole sample in Fig. 8a, which facilitates the nucleation and growth of crystal. In Fig. 8b, the G-I clusters are so rare that they are likely surrounded by other clusters like the G-II and G-III, and therefore they are separated from each other, so the crystallization is hindered and the formation of glass is therefore enhanced. The configuration marked by “ideal glass” represents a unique cluster that has the lowest energy among all the glassy clusters. However, the ideal glass still has a higher energy than its crystalline counterpart (i. e. the perfect crystal). It is expected that the ideal glass can be accessed at a cooling rate so slow that its entropy (e.g. Glass $a$) approaches that of the crystal in Fig. 6, but it has never been achieved in practice because the crystallization always intervenes. Logically, the ideal-glass cluster, involving only a microscopically tiny volume, must be available in a macroscopic sample. Perhaps, one can prepare ideal glass through transforming other clusters, e.g. G-I or G-II, into the ideal-glass one by annealing. In particular, G-II cluster is near the ideal-glass cluster as shown in Fig. 8, so the transformation likely happens to G-II cluster first. Of course, the energy barrier between them may be larger than the activation energy for crystallization, $\Delta G_x$ ($\sim k_B T_g$), so G-II clusters probably crystallize already before the transformation takes place during the annealing around $T_g$.

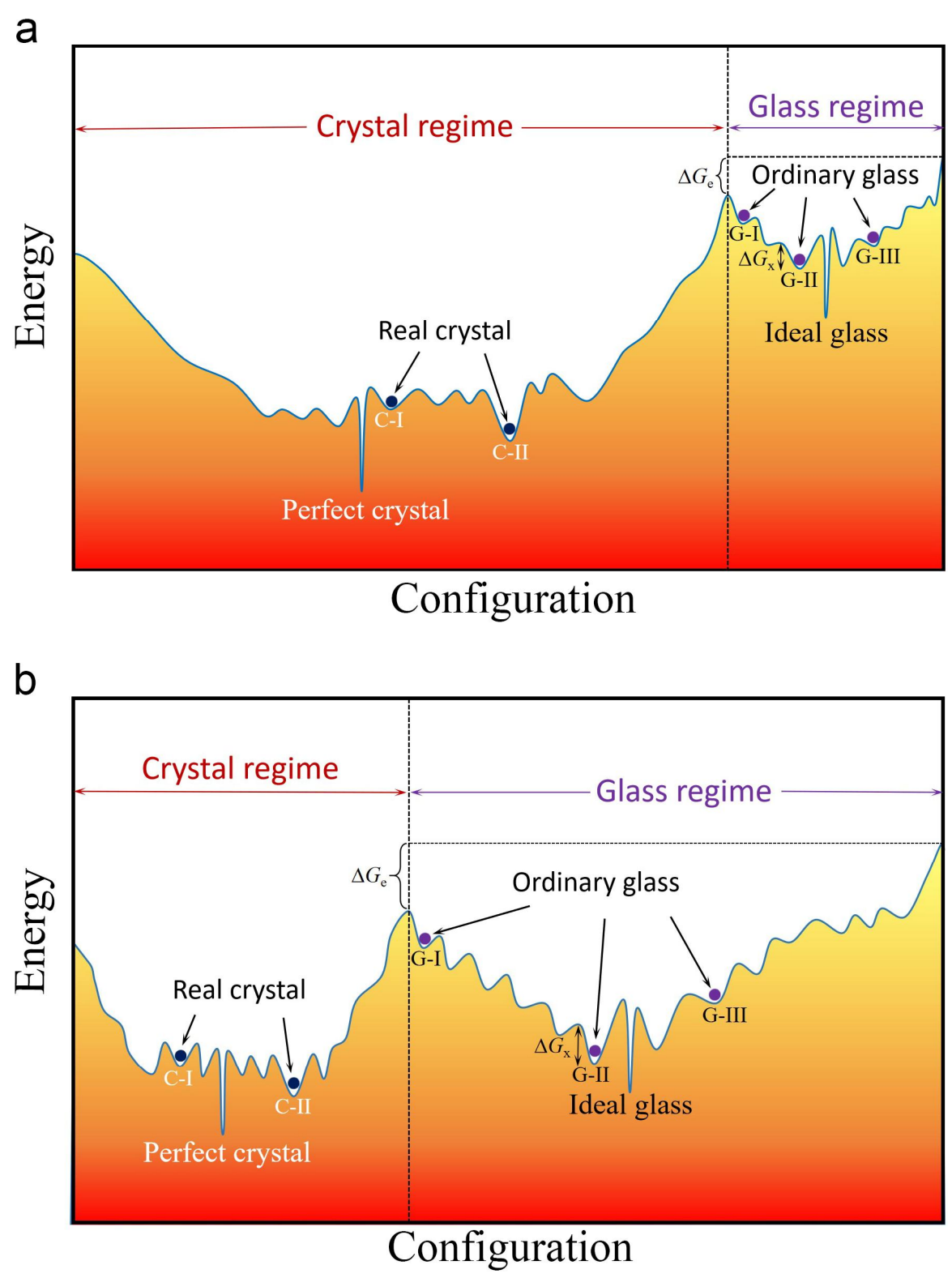

Fig. 8. 2D schematic of the energy landscape in a marginal glass-former (a) and a robust glass-former (b). The x-axis represents the possible configurations of clusters in the sample.

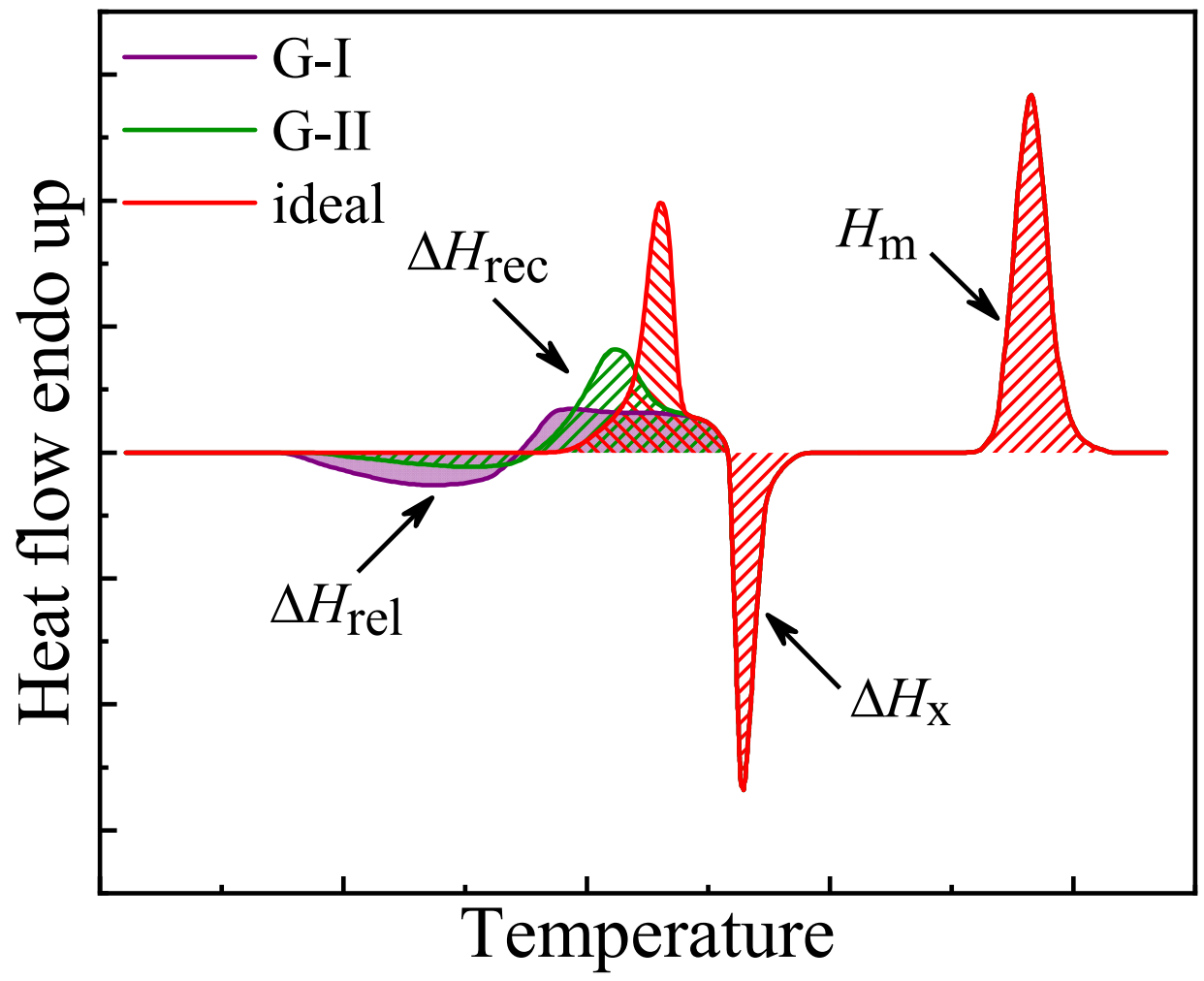

Fig. 9. An illustration of DSC trace in glass. G-I, G-II and ideal glasses are correspondingly marked in Fig.8. $\Delta H_{\mathrm{rel}}$, $\Delta H_{\mathrm{rec}}$, $\Delta H_{\mathrm{x}}$ and $\Delta H_{\mathrm{m}}$ are relaxation, recovery, crystallization and melting enthalpy, respectively. From the onset of crystallization, all the DSC traces are overlapped.

Annealing below $T_g$ usually induces structural relaxation, which releases energy, i.e. relaxation enthalpy $\Delta H_{\mathrm{rel}}$ shown by the DSC trace in Fig. 9. If G-I clusters dominate in number, the sample is called G-I glass. Similarly, we have G-II glass, G-III glass, ideal glass and so on. As shown in Fig. 9, $\Delta H_{\mathrm{rel}}$ for G-I glass has the maximum value (purple shaded area), and that for G-II glass is smaller (green pattern area), which reflects the energy difference between ordinary glasses and the ideal glass which has a zero $\Delta H_{\mathrm{rel}}$. On the other hand, the recovery enthalpy $\Delta H_{\mathrm{rec}}$ roughly presents the activation energy for a glass to return to SCL. It is illustrated as the height difference between the local basin where the glass (e.g. G-II) is trapped and the peak just located on the vertical dashed line which separates the crystal regime from the glass regime in Fig. 8. Undoubtedly, as presented in Fig. 9, $\Delta H_{\mathrm{rec}}$ for G-II glass (green pattern area) is greater than that for G-I glass (purple shaded area) but smaller than that for ideal glass (red pattern area) which actually has the maximum $\Delta H_{\mathrm{rec}}$. However, the crystallization enthalpy $\Delta H_{\mathrm{x}}$ is the same for all glasses (the red pattern area in Fig. 9), because the energy drops from the peak to a basin for the crystalline phase (e.g. C-II in Fig. 8) is almost a constant and it has nothing to do with the initial energy state of the corresponding glass before crystallization, as shown in Fig. 8.

## 7. A fundamental law in physics for the glass formation

In the above, the formation of metallic glass has been analyzed and understood in depth. But one may still occasionally fail to prepare a metallic glass with an average GFA. One of the reasons is that the crystalline precursor with a critical size develops coincidentally due to the composition and structure fluctuation and it grows rapidly in SCL during cooling process. It is worth noting that the crystalline precursor perhaps corresponds to a metastable crystalline phase which is absent in an equilibrium phase diagram such as Fig. 5. The appearance of metastable crystalline phase is mainly due to the nonequilibrium of quenching. This sometimes confuses the aforementioned empirical rule of the eutectic composition which, of course, is determined according to the equilibrium phase diagram. Another reason is, as Zallen had emphasized [1], that the melt cannot be cooled *fast enough* in physics. Greer and Sun [90] proposed that $T_g > T_m$ would be possible if the cooling rate were faster than $10^{13}$ K/s. However, there is no any literature on a metallic glass of $T_g > T_m$ in reality so far. It suggests that the cooling rate is fundamentally limited in a system subject to thermodynamic limit [91, 92]. In other words, no matter how hard one tries, the cooling rate is unable to increase all the time, around $T_g$ in particular. On the contrary, Johnson *et al.* [76] achieved a heating rate of ~$10^6$ K/s in processing MGs using capacitive discharge and prevented the intervening crystallization of MG in heating. If the pulse current is heavier in discharge, Ohmic heat will be more, and therefore the heating rate will be faster. In principle, the heating can be infinitely fast as long as the current is infinitely heavy. Unfortunately, there is no inverse Ohmic effect. The cooling is not just the reverse of the heating, and a great difference lies between their maximum rates.

Consider a closed system undergoing a heating or cooling process. Its temperature variation, $dT$, depends upon the transferred heat, $dQ$, then

$$dQ = C_p dT \tag{19}$$

where $C_p$ is the heat capacity. Undergoing an infinitesimal variation of temperature, $dT$, the process is reversible, so it has

$$dQ = TdS \tag{20}$$

Differentiating $Q$ with respect to the time $t$ gives

$$\frac{dT}{dt} = \frac{T}{C_{\mathrm{p}}}\frac{dS}{dt} \tag{21}$$

Based on the cooling and heating rates achieved in practice, a basic law is proposed for a closed system

$$\left(\frac{\partial S_{\mathrm{i}}}{\partial t}\right)_{\mathrm{max}} > \left(\frac{\partial S_{\mathrm{d}}}{\partial t}\right)_{\mathrm{max}} \tag{22}$$

where $S_{\mathrm{i}}$ and $S_{\mathrm{d}}$ respectively denote the increasing entropy and the decreasing entropy. The law in Eq. (22) demonstrates that the achievable maximum increasing-rate of entropy is greater than the achievable maximum decreasing-rate of entropy in a closed system. An isolated system can be treated as a special closed system, and its entropy will never decrease according to the second law of thermodynamics [93]. But one can claim that the entropy decreases negatively, i.e. $\frac{\partial S_{\mathrm{d}}}{\partial t} < 0$ , before the system reaches the equilibrium. In mathematics, it must exist: $\frac{\partial S_{\mathrm{d}}}{\partial t} = -\frac{\partial S_{\mathrm{i}}}{\partial t}$, then Eq. (22) can be written as

$$\left(\frac{\partial S_{\mathrm{i}}}{\partial t}\right)_{\mathrm{max}} > -\left(\frac{\partial S_{\mathrm{i}}}{\partial t}\right)_{\mathrm{max}} \quad \text{or} \quad \left(\frac{\partial S_{\mathrm{i}}}{\partial t}\right)_{\mathrm{max}} > 0 \tag{23}$$

which doesn't violate the second law of thermodynamics. In an isolated system under stationary equilibrium, the entropy will not change with time, so $\frac{\partial S_{\mathrm{i}}}{\partial t} = \frac{\partial S_{\mathrm{d}}}{\partial t} \equiv 0$. As a result, Eq. (22) must be modified as

$$\left(\frac{\partial S_{\mathrm{i}}}{\partial t}\right)_{\mathrm{max}} \geqslant \left(\frac{\partial S_{\mathrm{d}}}{\partial t}\right)_{\mathrm{max}} \tag{24}$$

when the isolated system is included. However, Eq. (23) is told by the second law of thermodynamics for an isolated system, so it is trivial in physics due to $\frac{\partial S_{\mathrm{i}}}{\partial t} > 0$. On the other hand, no system can be perfectly isolated in reality, so the interest to Eq. (24) is, if any, little. On the contrary, the law in Eq. (22) for closed systems has fundamental implications, especially for the formation of glass. It firmly ensures the accessibility of glass formation which is illustrated by the yellow line for glass in Fig. 6; otherwise, the yellow line will collapse to the cyan line for crystal, which keeps off the glass formation in a crystallizable material. Meanwhile, the law in Eq. (22) cannot be derived from other well-known laws of thermodynamics presented in textbook [93] or the controversial so-called fourth law of thermodynamics yet to be broadly recognized in the academic community [94-97]. Interestingly, the law proposed by Eq. (22) and the so-called fourth law of thermodynamics are both concerned with the changing rate of entropy in a system. In addition, the law in Eq. (22), like other well-known laws of thermodynamics [93, 98], could also be applied to any closed system of any subject such as ecology, economy and sociology provided the entropy is correctly formulated [93-97, 99].

**8. Phase transition in glass formation**

The issue of glass formation was first pointed out by Simon as early as 1930 [100, 101]. He held that the glass formation is a purely kinetic process, and the SCL and glass (the yellow line in Fig. 6) are in nonequilibrium state and therefore their entropy difference relative to crystalline counterpart (the cyan line in Fig. 6) will vanish if the cooling process is equilibrated [101]. Afterwards, the entropy of SCL was linearly extrapolated down to that of crystal at $T_K$ by Kauzmann [102], which leads to the entropy crisis or the famous Kauzmann paradox named by Angell [103]. The crisis can be averted via glass transition or crystallization according to Kauzmann's view on the thermal behavior of covalent glasses (e.g. $Si_2O_3$) and organic polymer glasses (e.g. polystyrene). These are good glass-formers that can hardly crystallize within the experimental time scale. As a result, Kauzmann alluded to the thermodynamic aspect of glass transition even though the kinetics was repeatedly stressed as a priority [102]. In fact, one must bear in mind that thermodynamic quantities such as entropy can be taken to describe the state of a system only if the system is in thermodynamic equilibrium and meets the requirement of thermodynamic limit [91, 92]. In such a system, there is neither metastable SCL nor glass if its stable crystalline phase with a lower energy exists. In Kauzmann's work [102], those SCLs were, at best, in thermal equilibrium with the surroundings, far from the thermodynamic equilibrium due to the very sluggish kinetics. Therefore, the extrapolation of entropy in SLC is meaningless in physics [104], and the entropy crisis seems like an imaginary quandary lacking of fundamental significance.

However, DiMarzio demonstrated that some materials like atactic polymer cannot crystallize at all during solidification subject to thermodynamic equilibrium [105, 106]. Perhaps, atactic polymer may be a practical example of ever glass mentioned above. Glass transition in ever glass is a thermodynamic process, albeit affected by the kinetic conditions applied. At this moment, the entropy, if extrapolated in Kauzmann's manner, will arrive at zero at a temperature, $T_K$, well above 0 K, as illustrated by the yellow dashed line in Fig.10. It certainly violates the third law of thermodynamics [107]. Glass transition, which turns the slope of entropy with respect to temperature towards the horizontal direction markedly at $T_g$, helps prevent this violation. Meanwhile, one can see that the entropy, $S$, is continuous with $T$, but it has a kink at $T_g$ on the yellow line in Fig. 10. In this case, the second order phase transition is expected according to Ehrenfest's classification scheme due to the discontinuity of the second derivative of Gibbs free energy, $G$, to $T$ herein, i.e. $\left(\frac{\partial^2 G}{\partial T^2}\right)_p = -\left(\frac{\partial S}{\partial T}\right)_p$ at $T_g$ [98]. This "discontinuity" has seemingly been observed in terms of isobaric specific heat $c_p = T\left(\frac{\partial S}{\partial T}\right)_p$ since 1922 [108-116]. Nevertheless, a lot of experimental measurements on various glassy matters show that $c_p$ is not stepwise discontinuous at $T_g$; instead, it undergoes a definitely continuous, although steep, jump when the glass transition takes place [101, 102, 108-110, 117]. On the other hand, a pronounced overshoot of $c_p$ often appears around $T_g$ on the heating curve in most glasses (refer to the peak for $\Delta H_{rec}$ in Fig. 9). $c_p$ seems to diverge at $T_g$. The divergence of $c_p$ characterizes the second order phase transition (i.e. $\lambda$-transition) for critical water, superfluid $^4$He and magnet at Curie temperature [93]. It corresponds to a singularity of $\frac{\partial S}{\partial T} \to +\infty$ but $S$ is unquestionably continuous, as shown by the green curve with a vertical tangent in Fig. 10. In fact, the $c_p$ overshoot is due to the relaxation towards the equilibrium for crystal and then the return back to the metastable SCL [118]. One can easily infer that there should be no such overshoot in ever glass during heating because the thermal process is in thermodynamic equilibrium within the whole temperature range. This trend can be observed in very robust glass former such as $Ge_2O$ in which the overshoot is almost imperceptible [54, 103]. Therefore, the overshoot is not an indicator of divergence for $c_p$ at $T_g$, and glass transition should not concern with second order phase transition, inconsistent with Gibbs and DiMarzio's proposal based on the configuration entropy [119].

The continuous and steep jump of $c_p$ at $T_g$ seems to indicate $\left.\frac{\partial c_p}{\partial T}\right|_{T=T_g} \to \infty$ , as shown by the cyan curve with a vertical tangent in Fig.S2 [120]. By definition, it has

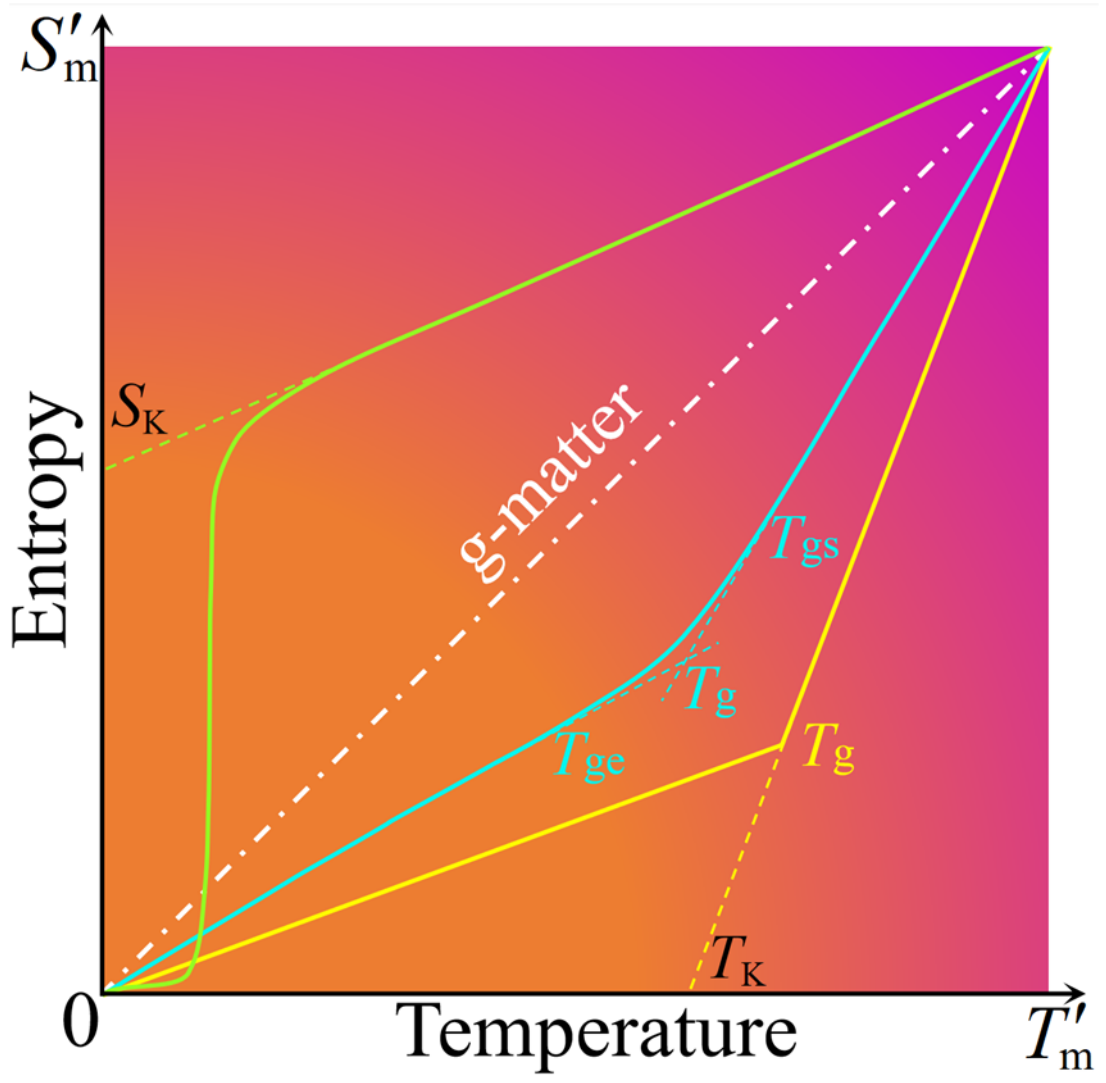

Fig.10. Temperature dependence of entropy in materials. $T_K$ is the Kauzmann temperature. $S_K$ denotes the residual entropy when $T = 0$ K. $S'_m$ is the entropy at $T'_m$, the temperature equivalent to $T_m$ in crystallizable material.

$$\left(\frac{\partial^3 G}{\partial T^3}\right)_p = -\left(\frac{\partial^2 S}{\partial T^2}\right)_p = \frac{c_p}{T^2} - \frac{1}{T}\frac{\partial c_p}{\partial T} \tag{25}$$

so the divergence of $\frac{\partial c_p}{\partial T}$ must lead to the divergence of $\left(\frac{\partial^3 G}{\partial T^3}\right)_p$. It demonstrates the third-order phase transition (TOPT) for glass formation, which was suggested by Woodcock according to the molecular dynamics simulation of hard spheres [121]. This supposal could be validated by measuring the $c_p$ of an ever glass such as atactic polymer. TOPT is rarely reported in the literature so far, and it was only observed in few cases, e.g. supercritical water by $c_p$ measurement and the x-ray small angle scattering in $CO_2$ and $CF_3H$ [122, 123]. Based on the Brownian motion, fourth-order phase transition for the glass formation subject to quasi-static cooling was proposed by Dimitrov [124], but there is no any substantial evidence in experiment yet. The glass transition usually occurs in a temperature interval between $T_{gs}$ and $T_{ge}$, as shown by the cyan curve in Fig.10. When $T_{gs}$ and $T_{ge}$ come towards each other and eventually coincide at $T_g$ where $S$ keeps differentiable, the cyan curve will approach the dash-dotted line which represents a unique matter (named as g-matter) with no any phase transition due to its infinitely differentiable $S$ with respect to $T$. In fact, the claim that there is no true phase transition for glass formation at all has been made many times; instead, the glass transition is attributed to an arrest of dynamics [125-128]. Meanwhile, during the course of approaching g-matter, fourth or even higher order phase transition may be possible. As the order of phase transition goes higher, the difference between neighboring phases is smeared [123], so the transition of greater than second-order is hard to describe and present in a clear and vivid way. Indeed, smeared phase transition has been introduced in disordered systems [129, 130]. Its basic scheme is relatively intuitive: Transition happens to different regions at different values of control parameter (e.g. temperature) due to the rugged energy landscape and irregular and aperiodic building blocks, very similar to that of random first-order transition (RFOT) [131]. In reality, the high order and smeared effect are likely coupled to dominate the glass formation, which profoundly darkens and obfuscates the physical image of glass transition. Interestingly, as illustrated in Fig.10, the green curve has a slope less than $\frac{S'_m}{T'_m}$ in high temperature range, which causes another entropy crisis: The entropy at 0 K, i.e. $S_K$, is greater than zero if an extrapolation regardless of underlying physics is conducted in Kauzmann's manner. Obviously, it goes against the third law of thermodynamics [107]. There must be a phase transition rather than a glass transition to avoid $S_K > 0$, which is signified by the vertical tangent on the green curve in Fig. 10. From Kauzmann's perspective, $B_2O_3$ may be a good example to manifest $S_K$ [102].

**9. Generalization of glass formation**

The issue of glass formation is challenging and long-standing. Almost one century has gone since it was first noticed in 1930 [108]. It led to the famous entropy crisis proposed by Kauzmann in 1948 [102], which still interests people nowadays [132]. Then, it was emphasized by Nobelist Philip Warren Anderson in 1995 [133], and was raised again in 2005 [134]. In 2008, debates on the nature of glass were even reported by New York Times [135], which attracted wide public attention. Principally, glass transition shares some attributes with traffic jamming and crowd-crushing accidents, so it can help enlighten the population and improve social management to some extent. However, no common agreement has not been reached on the issue, and people seem not to learn more definite knowledge nowadays than Kauzmann did over half a century ago. The lack of basic building blocks like primitive cell in crystals makes glasses notoriously difficult to be structurally distinguished from liquid, though they are entirely different in dynamics. Of course, "frozen-liquid" is not a satisfactory description of glass because it gives no any convincing details. The glass transition is also analyzed and explained by dynamic models [126, 127]. Ironically, dynamic features like heterogeneity in SCL are often attributed to the difference in static structure [136]. Fortunately, this dilemma does not hinder the development and application of glass too much. In particular, the synthesis of MGs, which must be implemented at a much higher cooling rate than that applied for polymer or oxide glass, highlights the nonequilibrium of the glass transition in thermodynamics. In other words, people should do their best to make the alloy deviate from thermodynamic equilibrium as far as possible to form an MG. One way is to reduce the number of atoms concerned (i. e. the size of sample, $R$, in Eq. (3)) so that the thermodynamic limit cannot be met [91, 92]. This has been unconsciously taken advantage of to prepare monatomic MGs in nanometer scale [14]. The other way is to shorten the time taken for the system to relax so that the equilibrium cannot be reached due to incomplete relaxation. In this sense, thermodynamic quantities that describe the state of an equilibrated system in textbooks are useless. There may be an unrecognized fundamental law to govern the glass formation, which is proposed as inequality (22). This law demonstrates the smaller decreasing-rate of entropy which makes the formation of glass available; otherwise, the excess entropy drops to zero instantly (see Fig.6) and the crystallization happens inevitably. MG is far from the ever glass, but its formation, as analyzed above, is, although difficult, possible to fulfill by virtue of inequality (22).

Indeed, no any MG has ever been discovered in nature and it is completely man-made instead. The relatively poor GFA, $\Psi_g$, in MGs forces people to improve the quenching technique, which in turn enhances $\Psi_g$. As aforementioned, there are a number of contributory factors, intrinsic (e. g. $\kappa$) and extrinsic (e. g. $T_{cm}$), for $\Psi_g$. Mathematically, it can be expressed as

$$\Psi_g = f(c_a, T_0, T_{cm}, p, \cdots) \quad (26)$$

where $c_a$ is the alloy composition. Obviously, none of these factors is able to determine $\Psi_g$ alone, and they must work together to figure out the best $\Psi_g$. In essence, it is a maximization or optimization problem of $\Psi_g$ in $n + 1$ dimensional space and $n$ is the number of factors in Eq. (26). At the same time, these factors are not completely independent, and they probably interact and even conflict. For instance, a high $T_0$ will, more or less, elevate $T_{cm}$ in practice, though $T_{cm}$ is supposed to keep constant in Eq. (2); a large $\eta$ will result in a small $D$ (see Eq.

(10)) which can lengthen $\tau_c$ and therefore enhance $\Psi_g$, but it lowers the fluidity and therefore mold-filling capacity of melt, and causes the cold lap on surface and porosity inside, then the heat conduction is blocked and $\Psi_g$ consequently deteriorates. In fact, this kind of multiobjective optimization problem has been scrutinized using the methodology and algorithm proposed by Pareto since 1906 [137]. The optimum $\Psi_g$ must locate in the Pareto front.

Based on the reported in the literature, the multi-component criterion proposed by Inoue and the confusion principle proposed by Greer seem applicable to most MGs, but they are not sufficient condition for inequality (5), which has been exemplified by a counterexample presented by Cantor *et al.* [138]; if other type of glass, e. g. elemental sulfur glass [139], is taken into account, they are not necessary condition either. Besides, that some binary alloys with positive mixing heat of constituent elements, e.g. Fe-Nb and U-V [60], have been quenched as glasses negates the necessity of the negative mixing heat of constituent elements in MG formers. These empirical rules including the preference of eutectic composition don't take every intrinsic factor in Eq. (9) into consideration, let alone the Eq. (3) for extrinsic factors. Hence, they are unable to certainly ensure inequality (5). In other words, one should choose extrinsic factors to accelerate $\dot{T}$ according to Eq. (3) and therefore decrease $\tau_g$, and select intrinsic factors to increase $\tau_c$ according to Eq. (9) to guarantee the inequality (5). In the ever-glass, the inequality (5) automatically holds because of, as suggested by Turnbull [16], $l \to \infty$, $\Delta G \to 0$ and therefore $\tau_c \to \infty$ in Eq. (9).

**10. Conclusion**

In summary, these formulations, calculations, analyses and generalizations on the glass formation are motivated by my own frequent and frustrating failures in MG preparation and the pioneers' seminal works, particularly Simon, Kauzmann, Turnbull and DiMarzio. The sufficient and necessary condition for the glass formation is emphasized in the inequality (5). The cooling process is analyzed by Eqs. (2) and (3) in a straightforward way, while crystallization is examined with the consideration of phase diagram in Eq. (9). It helps design robust glass formers and figure out the reason why sometimes there is a failure. The prerequisite for the well-known GFA criterion, $T_{rg}$, is reasonably deduced as $T_{cm} = 0$ K, which actually takes $T_{cm}$ into account for the first time. The phase transition, if really occurs, for glass formation is, at least, of the third order. Above all, a basic law of thermodynamics is proposed by the inequality (22), indicating the underlying physics for the glass formation: The decreasing rate of entropy is not favored in a closed system. This work tries to help understand the formation of glass technically and physically, particularly develop robust and new MGs. It must be reminded that occasional failure in preparing MG in which there usually exists a corresponding thermodynamically stable crystalline phase cannot be completely prevented even if Eq. (3) and Eq. (9) are both appreciated and optimized, because the crystallization might set in already just below $T_m$ due to a strong accidental fluctuation for which $\tau_c$ is infinitely close to zero in Eq. (9) so that the inequality (5) cannot be met any longer.

**Acknowledgment:** Support from Guangdong Major Basic and Applied Basic Research Foundation, China (Grant No. 2019B030302010), Guangdong Basic and Applied Basic Research Foundation, China (Grant No. 2020B1515120092) and National Natural Science Foundation of China (Grant Nos. 52071081 and 52271144) are all gratefully acknowledged. The author would like to thank Prof. Wang, Prof. Tong and Prof. Yang for their critical readings and kind suggestions.

Supplementary materials for "**On the formation of metallic glass**"

Wang Jian Guo

College of Mechanical Engineering, Dongguan University of Technology, Dongguan 523808, China

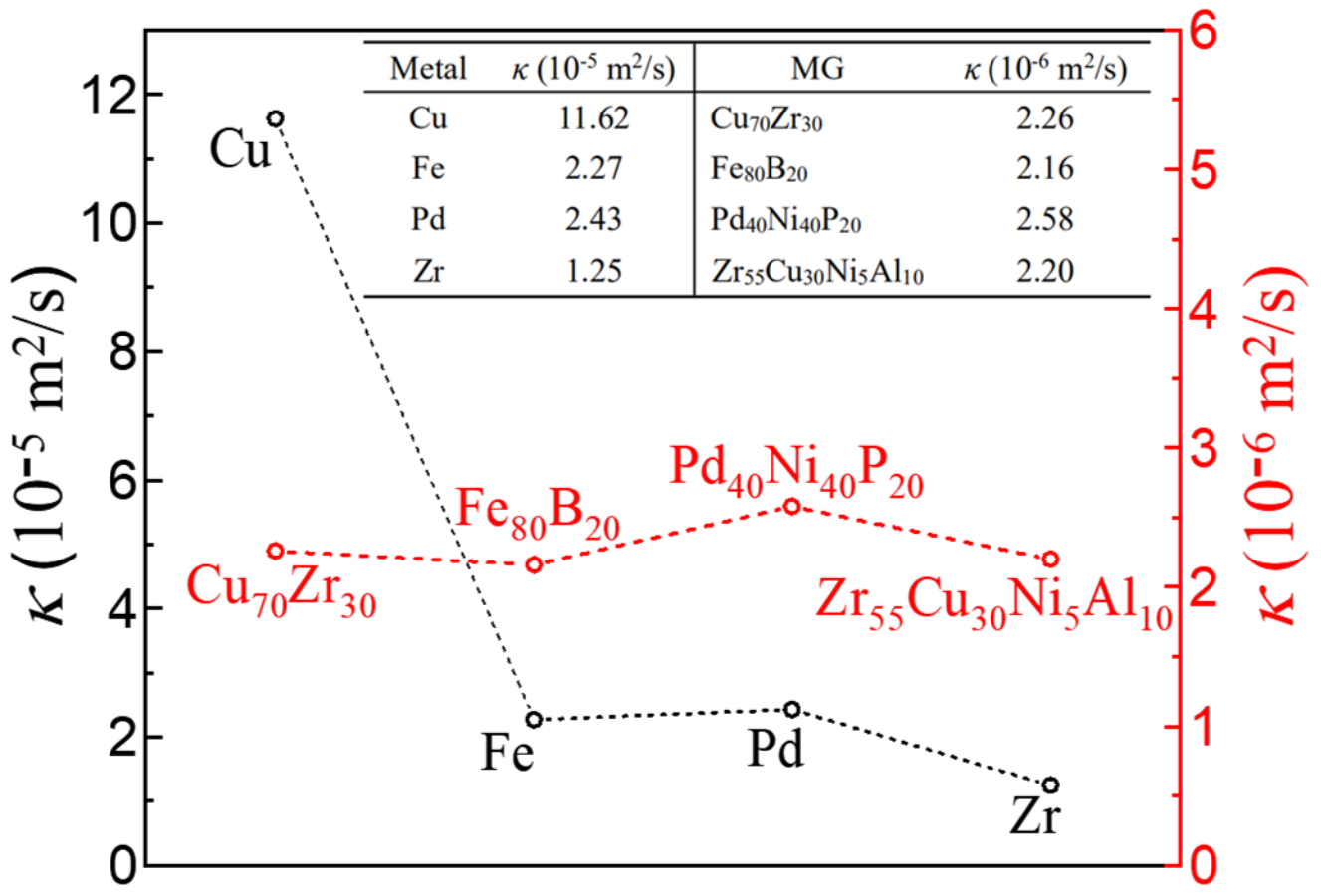

Fig.S1. The thermal diffusivity $\kappa$ in elemental metals (black) [1] and metallic glasses (red) [2, 3]. One can see that $\kappa$s for pure metals are significantly different and those for MGs are almost the same.

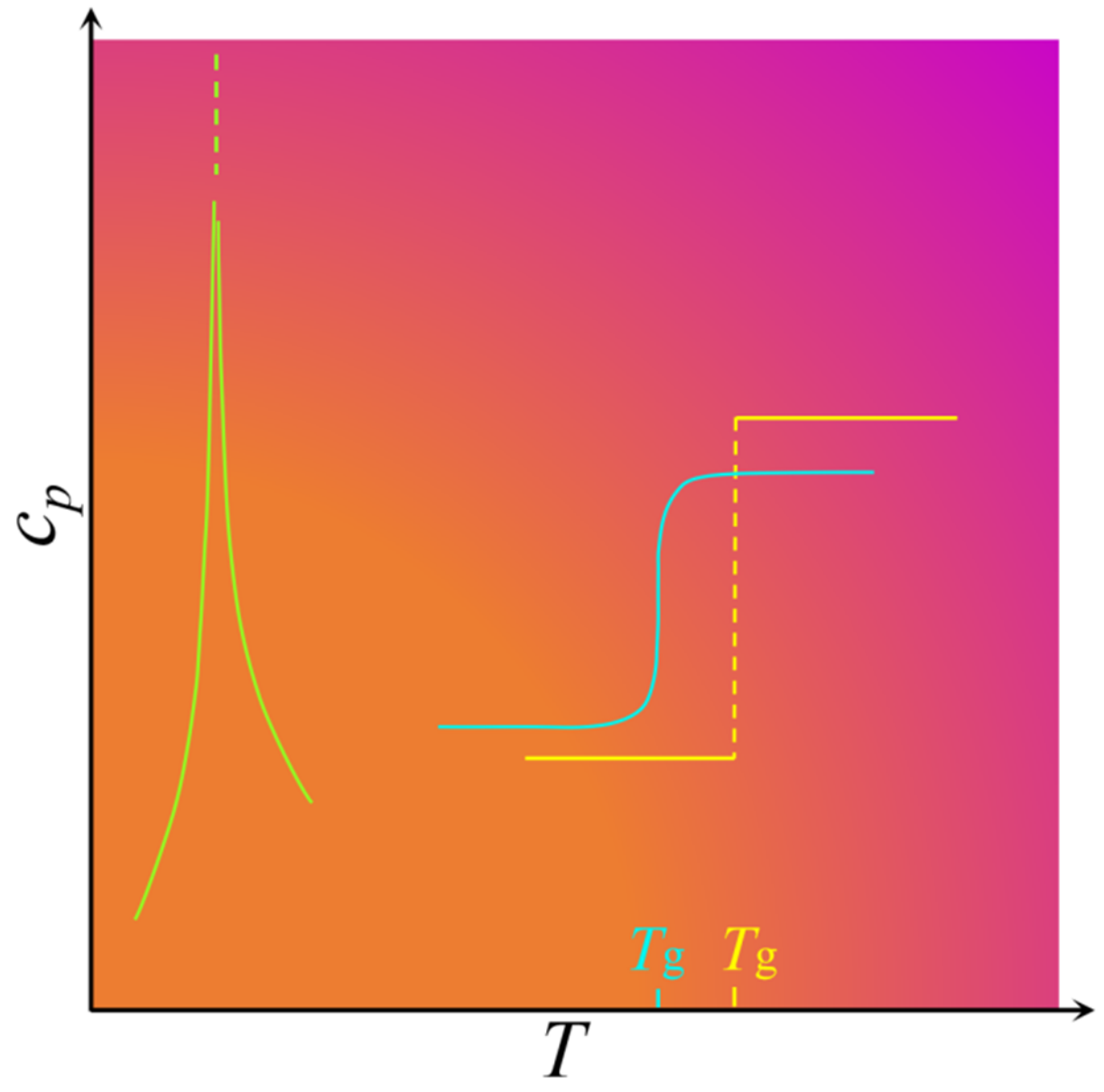

Fig.S2. The illustrated $c_p$ for different materials shown in Fig.12 of manuscript. The cyan curve has a vertical tangent on at $T_g$, which indicates the third order phase transition.